\documentclass[aps,prc,superscriptaddress,twoside,twocolumn,nofootinbib,10pt,%
showpacs,floatfix]{revtex4-1}

\usepackage{multirow}
\usepackage{amsmath,amssymb}
\usepackage{graphicx,bm}
\usepackage{slashed}
\usepackage{epstopdf}
\usepackage{ulem} 
\usepackage[usenames]{color}
\usepackage{float}
\usepackage{braket}
\usepackage{appendix}

\renewcommand\sout{\bgroup \color{red} \ULdepth=-.5ex \ULset}

\begin{document}

\title{Tribaryons in a constituent quark model}

\author{Aaron Park}
\email{aaron.park@yonsei.ac.kr}\affiliation{Department of Physics and Institute of Physics and Applied Physics, Yonsei University, Seoul 03722, Korea}
\author{Su Houng~Lee}\email{suhoung@yonsei.ac.kr}\affiliation{Department of Physics and Institute of Physics and Applied Physics, Yonsei University, Seoul 03722, Korea}
\date{\today}
\begin{abstract}
We  calculate the matrix elements of the color-spin interaction for all possible multi-quark states of tribaryons in flavor SU(3) broken case. For that purpose, we construct the flavor$\otimes$color$\otimes$spin wave functions of the  tribaryons, which are taken to be antisymmetric to satisfy the Pauli exclusion principle. Furthermore, we analyze the diquark structure of the tribaryon configurations using the symmetric and antisymmetric basis set of flavor, color and spin states.
\end{abstract}

\maketitle

\section{Introduction}
\label{Introduction}
While the interactions between quarks and gluons are described by QCD, which is based on color SU(3) gauge group, the phenomenological interaction between constituent quarks within a hadron can be taken to be mediated by gluons\cite{DeRujula:1975qlm} or mesons\cite{Glozman:1995fu}.   The constituent quark models can well reproduce the mass of normal hadrons, which are composed of three quarks for baryons and a quark-antiquark pair for mesons.
On the other hand, ever since Jaffe suggested the possibility of a stable dibaryon\cite{Jaffe:1976yi} and tetraquark using the bag model\cite{Jaffe:1976ig,Jaffe:1976ih}, exotic hadrons consisting of more than three quarks or quark antiquark have been a subject of experimental as well as theoretical studies(For a recent review, see \cite{Cho:2017dcy}).
The quark interaction within a multiquark configuration is the central theme in understanding exotic multi-quark configuration within the constituent quark model \cite{Liu:2019zoy}.

Multiquark configuration and their quark interactions can also be used to understand the baryon-baryon interaction as was first attempted successfully within the quark-cluster model\cite{Oka:1981ri}.  In fact, it was recently shown that the short distance part of the baryon-baryon interaction including those with strangeness extracted from the recent lattice calculation can be well reproduced using the constituent quark model\cite{Park:2019bsz}.  The crucial input in understanding the flavor dependence in these interactions is the color-spin interaction between quarks and the antisymmetrized flavor$\otimes$color$\otimes$spin wave functions that is required by the Pauli principle\cite{Oka:1981ri,Park:2019bsz}.

As for multiquark configuration involving three baryons, only a limited number of works exit \cite{Maezawa:2004va,Tamagaki:2008wf,Garcilazo:2016ylj,Park:2018ukx,Park:2019jff}.
The tribaryon configuration is related to the three nucleon interaction at short distance \cite{Park:2018ukx,Park:2019jff,Park:2019fko}, whose strength is important in relation to providing the stability of nuclear matter at high density.  For example,  the intrinsic three-body nuclear force including hyperons has been shown to provide a solution to the hyperon puzzle in a neutron star \cite{ Lattimer:2006xb,Bombaci:2016xzl}.  In a previous work, we have shown that the intrinsic three-body forces  vanish in the flavor SU(3) symmetric case, undermining the solution for the hyperon puzzle\cite{Park:2019jff}.  However, it is important to note that
the situation might change if SU(3) symmetry breaking effect is taken into account.
The importance of SU(3) breaking effect can not be overemphasized in the quark model.  While the color-spin interaction is large for the H-dibaryon in the SU(3) symmetric limit, the strength drastically reduces when the symmetry breaking is introduced\cite{Park:2016cmg}.

In this paper, we will calculate  the expectation values of the color-spin interaction of a tribaryon using a constituent quark model in the  flavor SU(3) breaking case. In particular, for each tribaryon configurations, we will explicitly calculate the
flavor$\otimes$color$\otimes$spin wave functions and analyze the diquark configurations in them.  This is accomplished by identifying the possible diquark configuration for all quark pairs within the tribaryon system and calculating the probability of each diquark configuration within the total wave function.
Through our study, we will elucidate the importance of flavor SU(3) breaking effect for each tribaryon configuration and highlight the important diquark configuration for attractive tribaryon configurations.

This paper is organized as follows. In Sec.\ref{Color-spin interaction}, we introduce the form of color-spin interaction and a parameter $\delta$ which can describe the flavor SU(3) breaking effect. In Sec.\ref{Wave function of a tribaryon}, we represent in detail  the method for constructing the wave function of a tribaryon. In Sec.\ref{Color-spin interaction matrix elements}, we show the color-spin matrix elements of the tribaryons for all possible quantum numbers in the flavor SU(3) broken case. In Sec.\ref{Diquark configuration of a tribaryon}, we analyze the diquark structures of the tribaryons using symmetric and antisymmetric basis set of flavor, color and spin wave function. Finally, Sec.\ref{Summary} is devoted to summary and concluding remarks.

\section{Color-spin interaction}
\label{Color-spin interaction}
In this work, we present the matrix elements of color-spin interaction for the tribaryons in the flavor SU(3) broken case. The color-spin interaction is defined  as follows.
\begin{align}
V_{CS}&=\sum_{i<j}^9 \frac{1}{m_i m_j}\lambda^c_i \lambda^c_j \sigma_i \cdot \sigma_j \nonumber \\
&\equiv \frac{1}{m_u m_u} H_{CS},
\label{color-spin}
\end{align}
where $\lambda^c_i$, $m_i$, $m_u$ are respectively the color SU(3) Gell-Mann matrices, the constituent quark mass of the $i$'th quark, and the constutient quark mass of $u,d$ quarks. In order to analyze the flavor SU(3) breaking effect, it is useful to introduce the following strange quark mass parameter.
\begin{align}
  \delta = 1-\frac{m_u}{m_s}.
\end{align}
When the mass of the strange quark becomes the same as that of the  $u,d$ quarks, the  flavor SU(3) breaking effect vanishes. For a totally antisymmetric flavor$\otimes$color$\otimes$spin wave function with a given flavor SU($N$), we can easily calculate the expectation value of the color-spin interaction using the following formula without considering the explicit form of the wavefunction.

\begin{widetext}
\begin{align}
   &-\sum_{i<j}^n \lambda^c_i \lambda^c_j \sigma_i \cdot \sigma_j = \frac{(N+6)}{3N}n^2 + (-2N-4)n + \frac{4}{3}S(S+1)+4C_F^{\mathrm{SU}(N)}+2C_C, \nonumber\\
  4C_F^{\mathrm{SU}(N)} &= \frac{2m}{N}(N^2-1)+\frac{4}{N}(N-1)\sum_{i=1}^{N-1} \binom{\sum_{j=1}^i p_{N-j}}{2} -4 \sum_{i=1}^{N-2}\left[\sum_{j=i+1}^{N-1}(j-i)p_j \left(1+\frac{1}{N}\sum_{k=i}^{N-1}p_k\right) \right],
\label{color-spin-SU(N)}
\end{align}
\end{widetext}
where $n$ is the total number of quarks, $C_F^{\mathrm{SU}(N)}=\frac{1}{4}\lambda^F \lambda^F$ and $C_C$ are the first kind of the Casimir Operator of flavor SU($N$) and color SU(3), respectively, $p_i$ is the number of columns containing $i$ boxes in a column in Young diagram, and $m$ is the total number of boxes minus $Np_N$, which is $\sum_{k=1}^{N-1}kp_k$.
For flavor SU(3) symmetric case,  Eq.(\ref{color-spin-SU(N)}) reduces to the following form.
\begin{small}
\begin{align}
  -\sum_{i<j}^n \lambda^c_i \lambda^c_j  \sigma_i \cdot \sigma_j &= n(n-10) + \frac{4}{3}S(S+1)+4C_F+2C_C, \nonumber\\
  4C_F &= \frac{4}{3}(p_1^2+p_2^2+3p_1+3p_2+p_1 p_2),
\end{align}
\end{small}
where $C_F$ is the first kind of the Casimir Operator of flavor SU(3). Then we can check that the matrix element of Eq.(\ref{color-spin-SU(N)}) becomes the same as the flavor SU(3) symmetric case when $N=3$.

\section{Wave function of a tribaryon}
\label{Wave function of a tribaryon}

In this work, we limit our discussion to the case where the orbital par of the quark wave function is totally symmetric.  This means that the
the remaining flavor$\otimes$color$\otimes$spin part of the wave function should be antisymmetric to satisfy the Pauli exclusion principle.
In general, if we consider the tribaryon as the sum of three baryon states, the orbital part does not have to be totally symmetric. We represent the possible orbital state of a tribaryon in Appendix \ref{orbital} for future work.

For the totally symmetric orbital state, there are eight possible flavor states for the tribaryons given as follows~\cite{Park:2018ukx}.\\

\begin{small}
$\begin{tabular}{|c|c|c|c}
  \cline{1-3}
  \quad \quad & \quad \quad & \quad \quad  \\
  \cline{1-3}
  \quad \quad & \quad \quad & \quad \quad \\
  \cline{1-3}
  \quad \quad & \quad \quad & \quad \quad \\
  \cline{1-3}
  \multicolumn{4}{c}{$\mathbf{1}(S=\frac{3}{2},\frac{5}{2},\frac{9}{2})$}
\end{tabular}$,
$\begin{tabular}{|c|c|c|c|c}
  \cline{1-4}
  \quad \quad & \quad \quad & \quad \quad & \quad \quad & \quad \quad  \\
  \cline{1-4}
  \quad \quad & \quad \quad & \quad \quad \\
  \cline{1-3}
  \quad \quad & \quad \quad \\
  \cline{1-2}
  \multicolumn{5}{l}{$\mathbf{8}(S=\frac{1}{2},\frac{3}{2},\frac{5}{2},\frac{7}{2})$}
\end{tabular}$,
$\begin{tabular}{|c|c|c|c|c|}
  \hline
  \quad \quad & \quad \quad & \quad \quad & \quad \quad & \quad \quad  \\
  \hline
  \quad \quad & \quad \quad  \\
  \cline{1-2}
  \quad \quad & \quad \quad \\
  \cline{1-2}
  \multicolumn{5}{c}{$\mathbf{10}(S=\frac{3}{2})$}
\end{tabular}$,
$\begin{tabular}{|c|c|c|c|}
  \hline
  \quad \quad & \quad \quad & \quad \quad & \quad \quad  \\
  \hline
  \quad \quad & \quad \quad & \quad \quad & \quad \quad \\
  \hline
  \quad \quad \\
  \cline{1-1}
  \multicolumn{4}{c}{$\mathbf{\bar{10}}(S=\frac{3}{2})$}
\end{tabular}$,
$\begin{tabular}{|c|c|c|c|c|}
  \hline
  \quad \quad & \quad \quad & \quad \quad & \quad \quad & \quad \quad  \\
  \hline
  \quad \quad & \quad \quad & \quad \quad \\
  \cline{1-3}
  \quad \quad \\
  \cline{1-1}
  \multicolumn{5}{c}{$\mathbf{27}(S=\frac{1}{2},\frac{3}{2},\frac{5}{2})$}
\end{tabular}$,
$\begin{tabular}{|c|c|c|c|c|c|}
  \hline
  \quad \quad & \quad \quad & \quad \quad & \quad \quad & \quad \quad & \quad \quad \\
  \hline
  \quad \quad & \quad \quad \\
  \cline{1-2}
  \quad \quad \\
  \cline{1-1}
  \multicolumn{6}{c}{$\mathbf{35}(S=\frac{1}{2})$}
\end{tabular}$,
$\begin{tabular}{|c|c|c|c|c|}
  \hline
  \quad \quad & \quad \quad & \quad \quad & \quad \quad & \quad \quad  \\
  \hline
  \quad \quad & \quad \quad & \quad \quad & \quad \quad \\
  \cline{1-4}
  \multicolumn{5}{c}{$\mathbf{\bar{35}}(S=\frac{1}{2})$}
\end{tabular}$,
$\begin{tabular}{|c|c|c|c|c|c|}
  \hline
  \quad \quad & \quad \quad & \quad \quad & \quad \quad & \quad \quad & \quad \quad \\
  \hline
  \quad \quad & \quad \quad & \quad \quad \\
  \cline{1-3}
  \multicolumn{6}{c}{$\mathbf{64}(S=\frac{3}{2})$ } \\
\end{tabular} $.
\end{small} \\

The brackets below each flavor state show the possible spin states. However, in flavor SU(3) broken case, the number of possible flavor states is limited in relation to isospin and strangeness.
As an example, for $I=2$,$S=\frac{3}{2}$ and one strange quark, only one SU(3) flavor state contributes
, which is flavor 64 multiplet.
\begin{small}
\begin{center}
$\begin{tabular}{|c|c|c|c|c|c|}
  \cline{1-6}
  \quad \quad & \quad \quad & \quad \quad & \quad \quad & \quad \quad & \quad \quad  \\
  \cline{1-6}
  \quad \quad & \quad \quad & $s$ \\
  \cline{1-3}
\end{tabular}_F$
\end{center}
\end{small}
Additionally, when there are more than one strange quark, the strange quarks  should not be located in the same column because their flavor wave function is symmetric. For example, the flavor 27 multiplet can accommodate the state with  $I=2$,$S=\frac{5}{2}$ and $strangeness=-3$.
\begin{small}
\begin{center}
$\begin{tabular}{|c|c|c|c|c|}
  \cline{1-5}
  \quad \quad & \quad \quad & \quad \quad & \quad \quad & \quad \quad  \\
  \cline{1-5}
  \quad \quad & $s$ & $s$ \\
  \cline{1-3}
  $s$ \\
  \cline{1-1}
\end{tabular}_F$
\end{center}
\end{small}
Similarly, the corresponding color$\otimes$spin coupling state can be constructed as follows.
\begin{small}
\begin{center}
$\begin{tabular}{|c|c|c|}
  \cline{1-3}
  \quad \quad & \quad \quad & \hspace{0.1cm}$s$\hspace{0.1cm}  \\
  \cline{1-3}
  \quad \quad & $s$   \\
  \cline{1-2}
  \quad \quad & $s$   \\
  \cline{1-2}
  \quad \quad \\
  \cline{1-1}
  \quad \quad \\
  \cline{1-1}
\end{tabular}_{CS}$
\end{center}
\end{small}
Now, using the symmetric property of Young-Yamanouchi basis~\cite{Stancu:1991rc}, we can represent the above flavor and color$\otimes$spin coupling state as follows.
\begin{widetext}
\begin{small}
  \begin{align}
    \Ket{\begin{tabular}{|c|c|c|c|c|}
    \cline{1-5}
    \quad \quad & \quad \quad & \quad \quad & \quad \quad & \quad \quad  \\
    \cline{1-5}
    \quad \quad & $s$ & $s$ \\
    \cline{1-3}
    $s$ \\
    \cline{1-1}
    \end{tabular}}_F =
    \frac{1}{3}~
    \Ket{\begin{tabular}{|c|c|c|c|c|}
    \cline{1-5}
    \quad \quad & \quad \quad & \quad \quad & \quad \quad & \quad \quad  \\
    \cline{1-5}
    \quad \quad & 7 & 8 \\
    \cline{1-3}
    9 \\
    \cline{1-1}
    \end{tabular}}_F
    +\frac{\sqrt{2}}{3}~
    \Ket{\begin{tabular}{|c|c|c|c|c|}
    \cline{1-5}
    \quad \quad & \quad \quad & \quad \quad & \quad \quad & \quad \quad  \\
    \cline{1-5}
    \quad \quad & 7 & 9 \\
    \cline{1-3}
    8 \\
    \cline{1-1}
    \end{tabular}}_F
    +\frac{\sqrt{2}}{\sqrt{3}}~
    \Ket{\begin{tabular}{|c|c|c|c|c|}
    \cline{1-5}
    \quad \quad & \quad \quad & \quad \quad & \quad \quad & \quad \quad  \\
    \cline{1-5}
    \quad \quad & 8 & 9 \\
    \cline{1-3}
    7 \\
    \cline{1-1}
    \end{tabular}}_F,\nonumber
  \end{align}
\end{small}
\begin{small}
  \begin{align}
    \Ket{\begin{tabular}{|c|c|c|}
    \cline{1-3}
    \quad \quad & \quad \quad & \hspace{0.1cm}$s$\hspace{0.1cm}  \\
    \cline{1-3}
    \quad \quad & $s$   \\
    \cline{1-2}
    \quad \quad & $s$   \\
    \cline{1-2}
    \quad \quad \\
    \cline{1-1}
    \quad \quad \\
    \cline{1-1}
    \end{tabular}}_{CS} =
    \frac{\sqrt{2}}{\sqrt{3}}~
    \Ket{\begin{tabular}{|c|c|c|}
    \cline{1-3}
    \quad \quad & \quad \quad & \hspace{0.1cm}7\hspace{0.1cm}  \\
    \cline{1-3}
    \quad \quad & 8   \\
    \cline{1-2}
    \quad \quad & 9   \\
    \cline{1-2}
    \quad \quad \\
    \cline{1-1}
    \quad \quad \\
    \cline{1-1}
    \end{tabular}}_{CS}
    -\frac{\sqrt{2}}{3}~
    \Ket{\begin{tabular}{|c|c|c|}
    \cline{1-3}
    \quad \quad & \quad \quad & \hspace{0.1cm}8\hspace{0.1cm}  \\
    \cline{1-3}
    \quad \quad & 7   \\
    \cline{1-2}
    \quad \quad & 9   \\
    \cline{1-2}
    \quad \quad \\
    \cline{1-1}
    \quad \quad \\
    \cline{1-1}
    \end{tabular}}_{CS}
    +\frac{1}{3}~
    \Ket{\begin{tabular}{|c|c|c|c|c|}
    \cline{1-3}
    \quad \quad & \quad \quad & \hspace{0.1cm}9\hspace{0.1cm}  \\
    \cline{1-3}
    \quad \quad & 7   \\
    \cline{1-2}
    \quad \quad & 8   \\
    \cline{1-2}
    \quad \quad \\
    \cline{1-1}
    \quad \quad \\
    \cline{1-1}
    \end{tabular}}_{CS}.
  \label{s-quark-symmetry}
  \end{align}
\end{small}
\end{widetext}
We can check that the above flavor and color$\otimes$spin coupling states are symmetric and antisymmetric under the exchange among 7,8 and 9, respectively. Now, we can construct the flavor$\otimes$color$\otimes$spin wave function as follows.
\begin{widetext}
\begin{small}
  \begin{align}
  \Ket{
    \left(\begin{tabular}{|c|}
      \cline{1-1}
      \hspace{0.1cm}$u$\hspace{0.1cm} \\
      \cline{1-1}
      \hspace{0.1cm}$u$\hspace{0.1cm} \\
      \cline{1-1}
      \hspace{0.1cm}$u$\hspace{0.1cm} \\
      \cline{1-1}
      \hspace{0.1cm}$u$\hspace{0.1cm} \\
      \cline{1-1}
      \hspace{0.1cm}$u$\hspace{0.1cm} \\
      \cline{1-1}
      \hspace{0.1cm}$u$\hspace{0.1cm} \\
      \cline{1-1}
      \end{tabular},
      \begin{tabular}{|c|}
      \cline{1-1}
      \hspace{0.1cm}$s$\hspace{0.1cm} \\
      \cline{1-1}
      \hspace{0.1cm}$s$\hspace{0.1cm} \\
      \cline{1-1}
      \hspace{0.1cm}$s$\hspace{0.1cm} \\
      \cline{1-1}
    \end{tabular}\right)}_{FCS}
    =\frac{1}{d_{[5,1]}^{\frac{1}{2}}}\sum_{i=1}^{d_{[5,1]}} (-1)^{n^f}_F
    \Ket{\begin{tabular}{|c|c|c|c|c|}
    \cline{1-5}
    $u_{1i}$ & $u_{2i}$ & $u_{3i}$ & $u_{4i}$ & $u_{5i}$  \\
    \cline{1-5}
    $u_{6i}$ & $s$ & $s$ \\
    \cline{1-3}
    $s$ \\
    \cline{1-1}
    \end{tabular}}_F \otimes
    \Ket{\begin{tabular}{|c|c|c|}
    \cline{1-3}
    $u_{1i}$ & $u_{6i}$ & \hspace{0.1cm}$s$\hspace{0.1cm}  \\
    \cline{1-3}
    $u_{2i}$ & $s$   \\
    \cline{1-2}
    $u_{3i}$ & $s$   \\
    \cline{1-2}
    $u_{4i}$ \\
    \cline{1-1}
    $u_{5i}$ \\
    \cline{1-1}
    \end{tabular}}_{CS}
  \end{align}
\end{small}
\end{widetext}
where $d_f$ is the dimension of the irreducible representation $f$ of $S_6$, $n^f_F$ is the number of transpositions necessary to bring the corresponding Young tableau of flavor state to a normal Young tableau~\cite{Stancu:1991rc} and $(u_{1i},u_{2i}...)$ is an array of numbers from 1 to 6 which can make the corresponding Young tableau.

Next, we need to decompose the color$\otimes$spin coupling state into color and spin states using the Clebsch-Gordan coefficients of $S_9$. Then the total wave function of a tribaryon with $I=2$,$S=\frac{5}{2}$ and $strnageness=-3$ can be represented as follows.

\begin{widetext}
\begin{small}
  \begin{align}
  \Psi_{FCS}=\Ket{
    \left(\begin{tabular}{|c|}
      \cline{1-1}
      \hspace{0.1cm}$u$\hspace{0.1cm} \\
      \cline{1-1}
      \hspace{0.1cm}$u$\hspace{0.1cm} \\
      \cline{1-1}
      \hspace{0.1cm}$u$\hspace{0.1cm} \\
      \cline{1-1}
      \hspace{0.1cm}$u$\hspace{0.1cm} \\
      \cline{1-1}
      \hspace{0.1cm}$u$\hspace{0.1cm} \\
      \cline{1-1}
      \hspace{0.1cm}$u$\hspace{0.1cm} \\
      \cline{1-1}
      \end{tabular},
      \begin{tabular}{|c|}
      \cline{1-1}
      \hspace{0.1cm}$s$\hspace{0.1cm} \\
      \cline{1-1}
      \hspace{0.1cm}$s$\hspace{0.1cm} \\
      \cline{1-1}
      \hspace{0.1cm}$s$\hspace{0.1cm} \\
      \cline{1-1}
    \end{tabular}\right)}_{FCS}
    =\sum_{F,C,S} a_{FCS}
    \Ket{\begin{tabular}{|c|c|c|c|c|}
    \cline{1-5}
    \quad \quad & \quad \quad & \quad \quad & \quad \quad & \quad \quad  \\
    \cline{1-5}
    \quad \quad & \quad \quad & \quad \quad \\
    \cline{1-3}
    \quad \quad \\
    \cline{1-1}
    \end{tabular}}_F \otimes
    \Ket{\begin{tabular}{|c|c|c|}
    \cline{1-3}
    \quad \quad & \quad \quad & \quad \quad  \\
    \cline{1-3}
    \quad \quad & \quad \quad & \quad \quad  \\
    \cline{1-3}
    \quad \quad & \quad \quad & \quad \quad  \\
    \cline{1-3}
    \end{tabular}}_C \otimes
    \Ket{\begin{tabular}{|c|c|c|c|c|c|c|}
    \cline{1-7}
    \quad \quad &\quad \quad &\quad \quad &\quad \quad &\quad \quad &\quad \quad &\quad \quad \\
    \cline{1-7}
    \quad \quad & \quad \quad \\
    \cline{1-2}
    \end{tabular}}_S,
  \end{align}
\end{small}
\end{widetext}
where $a_{FCS}$ are the coefficients which can be constructed from $n^F$,Clebsch-Gordan coefficients of $S_9$ and the coefficients in Eq.(\ref{s-quark-symmetry}). It should be noted that the blanks on the above equation are actually filled with numbers to form a Young tableau. After representing the wave function using each Young-Yamanouchi basis, we can easily calculate the matrix elements of color-spin interaction with the following formula.
\begin{align}
  \lambda^c_i \lambda^c_j = 2(ij)-\frac{2}{3}I, \nonumber \\
  \sigma_i\cdot \sigma_j = 2(ij)-I,
\end{align}
where $(ij)$ is the transposition between $i$-th and $j$-th quarks, and $I$ is the identity matrix in the corresponding irreducible representation. In general, the generator $\lambda$ of SU($N$) is as follows.
\begin{align}
  \lambda_i \lambda_j = 2(ij) -\frac{2}{N}.
\end{align}

\section{Color-spin interaction matrix elements}
\label{Color-spin interaction matrix elements}
Here, we represent the matrix elements of color-spin interaction for all possible tribaryon states in the flavor SU(3) broken case. In order to represent the symmetric property of the wave function, we use the $\{ijk\cdots\}$ notation which means antisymmetry under the exchange of particles among $i,j,k,\cdots$. For example, in the case of a tribaryon with three strange quarks, it should satisfy $\{123456\}\{789\}$ because we put the strange quarks on the last position of the wave function.

\subsection{$q^8 s:\{12345678\}9$}
\begin{itemize}
  \item $I=2, S=\frac{3}{2}(F=64)$ :\\
  $H_{CS}=56-\frac{16}{3}\delta$\\

  \item $I=2, S=\frac{1}{2}(F=35)$:\\
  $H_{CS}=40+\frac{32}{3}\delta$\\

  \item $I=1, S=\frac{5}{2}(F=27)$:\\
  $H_{CS}=\frac{104}{3}+\frac{16}{3}\delta$\\

  \item $I=1, S=\frac{3}{2}(F=64,27)$: \\
  $H_{CS}=\left(
  \begin{array}{cc}
  56-\frac{368}{21}\delta & \frac{16 \sqrt{10}}{21}\delta \\
  \frac{16 \sqrt{10}}{21}\delta & 28+\frac{172}{21}\delta \\
  \end{array}
  \right)$\\

  \item $I=1, S=\frac{1}{2}(F=\overline{35},27)$:\\
  $H_{CS}=\left(
  \begin{array}{cc}
  40-\frac{20}{3}\delta & -\frac{4}{3}\delta \\
  -\frac{4}{3}\delta & 24+\frac{28}{3}\delta \\
  \end{array}
  \right)$

  \item $I=0, S=\frac{3}{2}(F=\overline{10})$:\\
  $H_{CS}=20+\frac{20}{3}\delta$\\

  \item $I=0, S=\frac{1}{2}(F=\overline{35})$:\\
  $H_{CS}=40-\frac{40}{3}\delta$\\
\end{itemize}

\subsection{$q^7 s^2:\{1234567\}\{89\}$}
\begin{itemize}
  \item $I=\frac{5}{2}, S=\frac{3}{2}(F=64)$:\\
  $H_{CS}=56-\frac{32}{3}\delta+\frac{8}{3}\delta^2$\\

  \item $I=\frac{5}{2}, S=\frac{1}{2}(F=35)$:\\
  $H_{CS}=40+\frac{16}{3}\delta+\frac{8}{3}\delta^2$\\

  \item $I=\frac{3}{2}, S=\frac{5}{2}(F=27)$: \\
  $H_{CS}=\frac{104}{3}-\frac{16}{3}\delta+\frac{8}{3}\delta^2$\\

  \begin{widetext}
  \item $I=\frac{3}{2}, S=\frac{3}{2}(F=10,27,64)$:\\
  $H_{CS}=\left(
  \begin{array}{ccc}
  20+\frac{26}{3}\delta +\frac{26}{9}\delta^2 & -\frac{2\sqrt{35}}{3\sqrt{3}}\delta -\frac{2\sqrt{5}}{3\sqrt{21}}\delta^2 & \frac{4\sqrt{5}}{9\sqrt{7}}\delta^2 \\
  -\frac{2\sqrt{35}}{3\sqrt{3}}\delta -\frac{2\sqrt{5}}{3\sqrt{21}}\delta^2 & 28-\frac{2}{21}\delta +\frac{22}{7}\delta^2 & -\frac{80}{21\sqrt{3}}\delta -\frac{20}{21\sqrt{3}}\delta^2 \\
  \frac{4\sqrt{5}}{9\sqrt{7}}\delta^2 & -\frac{80}{21\sqrt{3}}\delta -\frac{20}{21\sqrt{3}}\delta^2 & 56-\frac{544}{21}\delta+\frac{208}{63}\delta^2 \\
  \end{array}
  \right)$

  \item $I=\frac{3}{2}, S=\frac{1}{2}(F=27,35,\overline{35})$:\\
  $H_{CS}=\left(
  \begin{array}{ccc}
  20+\frac{10}{3}\delta +3\delta^2 & -\frac{2}{3}\sqrt{\frac{5}{3}}\delta +\frac{1}{3}\sqrt{\frac{5}{3}}\delta^2 & \frac{8}{3\sqrt{3}}\delta +\frac{2}{3\sqrt{3}}\delta^2 \\
  -\frac{2}{3}\sqrt{\frac{5}{3}}\delta +\frac{1}{3}\sqrt{\frac{5}{3}}\delta^2 & 40-\frac{34}{3}\delta +\frac{29}{9}\delta^2 & \frac{2\sqrt{5}}{9}\delta^2 \\
  \frac{8}{3\sqrt{3}}\delta +\frac{2}{3\sqrt{3}}\delta^2 & \frac{2\sqrt{5}}{9}\delta^2 & 40-\frac{40}{3}\delta+\frac{28}{9}\delta^2 \\
  \end{array}
  \right)$
  \end{widetext}

  \item $I=\frac{1}{2}, S=\frac{7}{2}(F=8)$:\\
  $H_{CS}=24+\frac{8}{3}\delta^2$\\

  \item $I=\frac{1}{2}, S=\frac{5}{2}(F=27,8)$:\\
  $H_{CS}=\left(
  \begin{array}{cc}
  \frac{104}{3}-\frac{176}{15}\delta +\frac{8}{3}\delta^2 & -\frac{8}{5}\sqrt{\frac{7}{3}}\delta \\
  -\frac{8}{5}\sqrt{\frac{7}{3}}\delta & \frac{44}{3}+\frac{56}{15}\delta +\frac{8}{3}\delta^2 \\
  \end{array}
  \right)$

  \begin{widetext}
  \item $I=\frac{1}{2}, S=\frac{3}{2}(F=8,\overline{10},27,64)$:\\
  $H_{CS}=\left(
  \begin{array}{cccc}
  8+\frac{52}{5}\delta+\frac{10}{3}\delta^2 & -2\sqrt{\frac{2}{3}}\delta-\frac{1}{3}\sqrt{\frac{2}{3}}\delta^2 & \frac{14\sqrt{14}}{15}\delta +\frac{1}{3}\sqrt{\frac{2}{7}}\delta^2 & \frac{8}{3\sqrt{21}}\delta^2 \\
  -2\sqrt{\frac{2}{3}}\delta-\frac{1}{3}\sqrt{\frac{2}{3}}\delta^2 & 20-\frac{10}{3}\delta+\frac{25}{9}\delta^2 & \frac{2}{3}\sqrt{\frac{7}{3}}\delta -\frac{1}{3\sqrt{21}}\delta^2 & -\frac{4}{9}\sqrt{\frac{2}{7}}\delta^2 \\
  \frac{14\sqrt{14}}{15}\delta +\frac{1}{3}\sqrt{\frac{2}{7}}\delta^2 & \frac{2}{3}\sqrt{\frac{7}{3}}\delta -\frac{1}{3\sqrt{21}}\delta^2 & 28-\frac{982}{105}\delta +\frac{19}{7}\delta^2 & -\frac{80}{21}\sqrt{\frac{2}{3}}\delta +\frac{4}{21}\sqrt{\frac{2}{3}}\delta^2 \\
  \frac{8}{3\sqrt{21}}\delta^2 & -\frac{4}{9}\sqrt{\frac{2}{7}}\delta^2 & -\frac{80}{21}\sqrt{\frac{2}{3}}\delta +\frac{4}{21}\sqrt{\frac{2}{3}}\delta^2 & 56-\frac{736}{21}\delta+\frac{200}{63}\delta^2 \\
  \end{array}
  \right)$

  \item $I=\frac{1}{2}, S=\frac{1}{2}(F=8,27,\overline{35})$: \\
  $H_{CS}=\left(
  \begin{array}{ccc}
  4+\frac{32}{3}\delta+\frac{28}{9}\delta^2 & -\frac{8}{3\sqrt{3}}\delta-\frac{2}{3\sqrt{3}}\delta^2 & -\frac{2\sqrt{5}}{9}\delta^2 \\
  -\frac{8}{3\sqrt{3}}\delta-\frac{2}{3\sqrt{3}}\delta^2 & 24-\frac{26}{3}\delta+3\delta^2 & -\frac{2}{3}\sqrt{\frac{5}{3}}\delta +\frac{1}{3}\sqrt{\frac{5}{3}}\delta^2 \\
  -\frac{2\sqrt{5}}{9}\delta^2 & -\frac{2}{3}\sqrt{\frac{5}{3}}\delta +\frac{1}{3}\sqrt{\frac{5}{3}}\delta^2 & 40-\frac{70}{3}\delta+\frac{29}{9}\delta^2 \\
  \end{array}
  \right)$
  \end{widetext}
\end{itemize}

\subsection{$q^6 s^3:\{123456\}\{789\}$}
\begin{itemize}
  \item $I=3, S=\frac{3}{2}(F=64)$:\\
  $H_{CS}=56-16\delta+8\delta^2$\\

  \item $I=2, S=\frac{5}{2}(F=27)$:\\
  $H_{CS}=\frac{104}{3}-16\delta+8\delta^2$\\

  \item $I=2, S=\frac{3}{2}(F=27,64)$:\\
  $H_{CS}=\left(
  \begin{array}{cc}
  28-\frac{176}{21}\delta +\frac{178}{21}\delta^2 & \frac{16\sqrt{5}}{21}\delta +\frac{8\sqrt{5}}{21}\delta^2 \\
  \frac{16\sqrt{5}}{21}\delta +\frac{8\sqrt{5}}{21}\delta^2 & 56-\frac{240}{7}\delta +\frac{200}{21}\delta^2 \\
  \end{array}
  \right)$

  \begin{widetext}
  \item $I=2, S=\frac{1}{2}(F=27,35,\overline{35})$:\\
  $H_{CS}=\left(
  \begin{array}{ccc}
  24-\frac{8}{3}\delta+\frac{28}{3}\delta^2 & -\frac{4}{3}\delta-\frac{2}{3}\delta^2 & -\frac{4}{3}\delta-\frac{2}{3}\delta^2 \\
  -\frac{4}{3}\delta-\frac{2}{3}\delta^2 & 40-20\delta+\frac{28}{3}\delta^2 & -\frac{2}{3}\delta^2 \\
  -\frac{4}{3}\delta-\frac{2}{3}\delta^2 & -\frac{2}{3}\delta^2 & 40-20\delta+\frac{28}{3}\delta^2 \\
  \end{array}
  \right)$
  \end{widetext}

  \item $I=1, S=\frac{7}{2}(F=8)$:\\
  $H_{CS}=24-16\delta+8\delta^2$\\

  \item $I=1, S=\frac{5}{2}(F=8,27)$:\\
  $H_{CS}=\left(
  \begin{array}{cc}
  \frac{44}{3}-\frac{24}{5}\delta+\frac{134}{15}\delta^2 & \frac{8\sqrt{14}}{15}\delta+\frac{4\sqrt{14}}{15}\delta^2 \\
  \frac{8\sqrt{14}}{15}\delta+\frac{4\sqrt{14}}{15}\delta^2 & \frac{104}{3}-\frac{368}{15}\delta+\frac{136}{15}\delta^2 \\
  \end{array}
  \right)$

  \begin{widetext}
  \item $I=1, S=\frac{3}{2}(F=8,10,\overline{10},27,64)$:\\
  $H_{CS}=\left(
  \begin{array}{ccccc}
  8-\frac{24}{5}\delta+\frac{392}{45}\delta^2 & -2\sqrt{\frac{2}{3}}\delta-\frac{1}{9}\sqrt{\frac{2}{3}}\delta^2 & 2\sqrt{\frac{2}{3}}\delta+\frac{1}{9}\sqrt{\frac{2}{3}}\delta^2 & -\frac{28}{15}\sqrt{\frac{7}{3}}\delta-\frac{58}{15\sqrt{21}}\delta^2 & \frac{8}{9}\sqrt{\frac{5}{21}}\delta^2 \\
  -2\sqrt{\frac{2}{3}}\delta-\frac{1}{9}\sqrt{\frac{2}{3}}\delta^2 & 20-\frac{40}{3}\delta+\frac{245}{27}\delta^2 & \frac{25}{27}\delta^2 & -\frac{4\sqrt{14}}{9}\delta -\frac{1}{9}\sqrt{\frac{2}{7}}\delta^2 & -\frac{8}{27}\sqrt{\frac{10}{7}}\delta^2 \\
  2\sqrt{\frac{2}{3}}\delta +\frac{1}{9}\sqrt{\frac{2}{3}}\delta^2 & \frac{25}{27}\delta^2 & 20-\frac{40}{3}\delta+\frac{245}{27}\delta^2 & \frac{4\sqrt{14}}{9}\delta+\frac{1}{9}\sqrt{\frac{2}{7}}\delta^2 & \frac{8}{27}\sqrt{\frac{10}{7}}\delta^2 \\
  -\frac{28}{15}\sqrt{\frac{7}{3}}\delta-\frac{58}{15\sqrt{21}}\delta^2 & -\frac{4\sqrt{14}}{9}\delta-\frac{1}{9}\sqrt{\frac{2}{7}}\delta^2 & \frac{4\sqrt{14}}{9}\delta+\frac{1}{9}\sqrt{\frac{2}{7}}\delta^2 & 28-\frac{2176}{105}\delta+\frac{992}{105}\delta^2 & \frac{80\sqrt{5}}{63}\delta+\frac{8\sqrt{5}}{63}\delta^2 \\
  \frac{8}{9}\sqrt{\frac{5}{21}}\delta^2 & -\frac{8}{27}\sqrt{\frac{10}{7}}\delta^2 & \frac{8}{27}\sqrt{\frac{10}{7}}\delta^2 & \frac{80\sqrt{5}}{63}\delta+\frac{8\sqrt{5}}{63}\delta^2 & 56-\frac{976}{21}\delta+\frac{1832}{189}\delta^2 \\
  \end{array}
  \right)$

  \item $I=1, S=\frac{1}{2}(F=8,27,35,\overline{35})$:\\
  $H_{CS}=\left(
  \begin{array}{cccc}
  4+\frac{8}{3}\delta+\frac{254}{27}\delta^2 & \frac{8\sqrt{2}}{9}\delta-\frac{4\sqrt{2}}{9}\delta^2 & -\frac{4\sqrt{10}}{27}\delta^2 & -\frac{4\sqrt{10}}{27}\delta^2 \\
  \frac{8\sqrt{2}}{9}\delta-\frac{4\sqrt{2}}{9}\delta^2 & 24-\frac{56}{3}\delta+\frac{28}{3}\delta^2 & \frac{4\sqrt{5}}{9}\delta-\frac{2\sqrt{5}}{9}\delta^2 & \frac{4\sqrt{5}}{9}\delta-\frac{2\sqrt{5}}{9}\delta^2 \\
  -\frac{4\sqrt{10}}{27}\delta^2 & \frac{4\sqrt{5}}{9}\delta-\frac{2\sqrt{5}}{9}\delta^2 & 40-\frac{100}{3}\delta+\frac{260}{27}\delta^2 & -\frac{10}{27}\delta^2 \\
  -\frac{4\sqrt{10}}{27}\delta^2 & \frac{4\sqrt{5}}{9}\delta-\frac{2\sqrt{5}}{9}\delta^2 & -\frac{10}{27}\delta^2 & 40-\frac{100}{3}\delta+\frac{260}{27}\delta^2 \\
  \end{array}
  \right)$
  \end{widetext}

  \item $I=0, S=\frac{9}{2}(F=1)$:\\
  $H_{CS}=24-16\delta+8\delta^2$\\

  \item $I=0, S=\frac{7}{2}(F=8)$:\\
  $H_{CS}=24-16\delta+8\delta^2$\\

  \begin{widetext}
  \item $I=0, S=\frac{5}{2}(F=1,8,27)$:\\
  $H_{CS}=\left(
  \begin{array}{ccc}
  \frac{8}{3}-\frac{16  }{9}\delta+\frac{80}{9}\delta ^2 & -\frac{8 \sqrt{35} }{9}\delta-\frac{4}{9} \sqrt{\frac{7}{5}} \delta ^2 & \frac{4}{3} \sqrt{\frac{2}{5}} \delta ^2 \\
  -\frac{8 \sqrt{35} }{9}\delta -\frac{4}{9} \sqrt{\frac{7}{5}} \delta ^2 & \frac{44}{3} -\frac{664}{45}\delta+\frac{374 }{45}\delta ^2 & \frac{4 \sqrt{14}}{5}\delta-\frac{2 \sqrt{14}}{15} \delta ^2 \\
  \frac{4}{3} \sqrt{\frac{2}{5}} \delta ^2 & \frac{4 \sqrt{14}}{5}\delta-\frac{2 \sqrt{14}}{15}\delta ^2 & \frac{104}{3} -\frac{144}{5}\delta+\frac{44}{5}\delta ^2 \\
  \end{array}
  \right)$

  \item $I=0, S=\frac{3}{2}(F=1,8,27,64)$:\\
  $H_{CS}=\left(
  \begin{array}{cccc}
  -4 +\frac{8}{3}\delta+\frac{55}{6}\delta ^2 & -\frac{2 \sqrt{35}}{3}\delta -\frac{1}{3} \sqrt{\frac{7}{5}} \delta ^2&
   -\frac{7}{2 \sqrt{15}}\delta ^2 & 0 \\
  -\frac{2 \sqrt{35}}{3}\delta -\frac{1}{3} \sqrt{\frac{7}{5}} \delta ^2 & 8-\frac{88}{15} \delta+\frac{128}{15}\delta ^2
   & -\frac{14}{5} \sqrt{\frac{7}{3}} \delta +\frac{1}{5} \sqrt{\frac{3}{7}} \delta ^2 & \frac{8}{\sqrt{105}}\delta ^2 \\
  -\frac{7}{2 \sqrt{15}}\delta ^2 & -\frac{14}{5} \sqrt{\frac{7}{3}} \delta +\frac{1}{5} \sqrt{\frac{3}{7}} \delta ^2 &
   28-\frac{2824}{105}\delta+\frac{1843}{210}\delta ^2 & \frac{32 \sqrt{5}}{21}\delta-\frac{16}{21 \sqrt{5}}\delta ^2
   \\
  0 & \frac{8}{\sqrt{105}}\delta ^2 & \frac{32 \sqrt{5}}{21}\delta-\frac{16}{21 \sqrt{5}}\delta ^2 & 56-\frac{368 }{7}\delta+\frac{200}{21} \delta^2 \\
  \end{array}
  \right)$
  \end{widetext}

  \item $I=0, S=\frac{1}{2}(F=8,27)$:\\
  $H_{CS}=\left(
  \begin{array}{cc}
  4-8 \delta +\frac{26}{3}\delta ^2 & \frac{4 \sqrt{2}}{3}\delta +\frac{2 \sqrt{2}}{3}\delta ^2 \\
  \frac{4 \sqrt{2}}{3}\delta +\frac{2 \sqrt{2}}{3}\delta ^2 & 24-\frac{80}{3}\delta+\frac{28}{3}\delta ^2 \\
  \end{array}
  \right)$

\end{itemize}

\subsection{$q^5 s^4:\{12345\}\{6789\}$}
\begin{itemize}
  \item $I=\frac{5}{2}, S=\frac{3}{2}(F=64)$:\\
  $H_{CS}=56-\frac{128}{3}\delta+\frac{56}{3}\delta^2$\\

  \item $I=\frac{5}{2}, S=\frac{1}{2}(F=\overline{35})$:\\
  $H_{CS}=40-\frac{80}{3}\delta+\frac{56}{3}\delta^2$\\

  \item $I=\frac{3}{2}, S=\frac{5}{2}(F=27)$:\\
  $H_{CS}=\frac{104}{3}-\frac{112}{3}\delta+\frac{56}{3}\delta^2$\\

  \begin{widetext}
  \item $I=\frac{3}{2}, S=\frac{3}{2}(F=\overline{10},27,64)$:\\
  $H_{CS}=\left(
  \begin{array}{ccc}
  20-\frac{70}{3}\delta+\frac{170}{9}\delta ^2 & \frac{2}{3} \sqrt{\frac{35}{3}} \delta +\frac{2}{3} \sqrt{\frac{5}{21}}
   \delta ^2 & -\frac{4}{9} \sqrt{\frac{5}{7}} \delta ^2 \\
  \frac{2}{3} \sqrt{\frac{35}{3}} \delta +\frac{2}{3} \sqrt{\frac{5}{21}} \delta ^2 & 28-\frac{674}{21}\delta+\frac{134
   }{7}\delta ^2 & -\frac{80}{21 \sqrt{3}}\delta-\frac{20}{21 \sqrt{3}} \delta ^2 \\
  -\frac{4}{9} \sqrt{\frac{5}{7}} \delta ^2 & -\frac{80}{21 \sqrt{3}}\delta-\frac{20}{21 \sqrt{3}}\delta ^2 &
   56-\frac{1216}{21}\delta+\frac{1216}{63}\delta ^2 \\
  \end{array}
  \right)$

  \item $I=\frac{3}{2}, S=\frac{1}{2}(F=27,35,\overline{35})$:\\
  $H_{CS}=\left(
  \begin{array}{ccc}
  24-\frac{86}{3}\delta+19 \delta ^2 & \frac{8}{3 \sqrt{3}}\delta+\frac{2}{3 \sqrt{3}}\delta ^2 & -\frac{2}{3}
   \sqrt{\frac{5}{3}} \delta +\frac{1}{3} \sqrt{\frac{5}{3}} \delta ^2 \\
  \frac{8}{3 \sqrt{3}}\delta+\frac{2}{3 \sqrt{3}}\delta ^2 & 40-\frac{136}{3}\delta+\frac{172}{9}\delta ^2 & \frac{2
   \sqrt{5}}{9}\delta ^2 \\
  -\frac{2}{3} \sqrt{\frac{5}{3}} \delta +\frac{1}{3} \sqrt{\frac{5}{3}} \delta ^2 & \frac{2 \sqrt{5}}{9}\delta ^2 &
   40-\frac{130}{3}\delta+\frac{173}{9}\delta ^2 \\
  \end{array}
  \right)$
  \end{widetext}

  \item $I=\frac{1}{2}, S=\frac{7}{2}(F=8)$:\\
  $H_{CS}=24-32\delta+\frac{56}{3}\delta^2$\\

  \item $I=\frac{1}{2}, S=\frac{5}{2}(F=8,27)$:\\
  $H_{CS}=\left(
  \begin{array}{cc}
  \frac{44}{3}-\frac{424}{15}\delta+\frac{56}{3}\delta ^2 & -\frac{8}{5} \sqrt{\frac{7}{3}} \delta  \\
  -\frac{8}{5} \sqrt{\frac{7}{3}} \delta  & \frac{104}{3}-\frac{656 }{15}\delta+\frac{56}{3}\delta ^2 \\
  \end{array}
  \right)$

  \begin{widetext}
  \item $I=\frac{1}{2}, S=\frac{3}{2}(F=8,10,27,64)$:\\
  $H_{CS}=\left(
  \begin{array}{cccc}
  8-\frac{108}{5}\delta+\frac{58}{3}\delta ^2 & 2 \sqrt{\frac{2}{3}} \delta +\frac{1}{3} \sqrt{\frac{2}{3}} \delta ^2 &
   \frac{14 \sqrt{14}}{15}\delta+\frac{1}{3} \sqrt{\frac{2}{7}} \delta ^2 & \frac{8}{3 \sqrt{21}}\delta ^2 \\
  2 \sqrt{\frac{2}{3}} \delta +\frac{1}{3} \sqrt{\frac{2}{3}} \delta ^2 & 20-\frac{106}{3}\delta+\frac{169}{9}\delta ^2 &
   -\frac{2}{3} \sqrt{\frac{7}{3}} \delta +\frac{1}{3 \sqrt{21}}\delta ^2 & \frac{4}{9} \sqrt{\frac{2}{7}} \delta ^2 \\
  \frac{14 \sqrt{14}}{15}\delta+\frac{1}{3} \sqrt{\frac{2}{7}} \delta ^2 & -\frac{2}{3} \sqrt{\frac{7}{3}} \delta
   +\frac{1}{3 \sqrt{21}}\delta ^2 & 28-\frac{4342}{105}\delta+\frac{131 }{7}\delta ^2 & -\frac{80}{21}
   \sqrt{\frac{2}{3}} \delta +\frac{4}{21} \sqrt{\frac{2}{3}} \delta ^2 \\
  \frac{8}{3 \sqrt{21}}\delta ^2 & \frac{4}{9} \sqrt{\frac{2}{7}} \delta ^2 & -\frac{80}{21}  \sqrt{\frac{2}{3}} \delta
   +\frac{4}{21} \sqrt{\frac{2}{3}} \delta ^2 & 56-\frac{1408 }{21}\delta+\frac{1208}{63}\delta ^2 \\
  \end{array}
  \right)$

  \item $I=\frac{1}{2}, S=\frac{1}{2}(F=8,27,35)$:\\
  $H_{CS}=\left(
  \begin{array}{ccc}
  4-\frac{64}{3}\delta+\frac{172}{9}\delta ^2 & -\frac{8}{3 \sqrt{3}}\delta-\frac{2}{3 \sqrt{3}}\delta ^2 & -\frac{2}{9}
   \sqrt{5} \delta ^2 \\
  -\frac{8}{3 \sqrt{3}}\delta-\frac{2}{3 \sqrt{3}}\delta ^2 & 24-\frac{122}{3}\delta+19 \delta ^2 & -\frac{2}{3}
   \sqrt{\frac{5}{3}} \delta +\frac{1}{3} \sqrt{\frac{5}{3}} \delta ^2 \\
  \frac{1}{9} (-2) \sqrt{5} \delta ^2 & -\frac{2}{3} \sqrt{\frac{5}{3}} \delta +\frac{1}{3} \sqrt{\frac{5}{3}} \delta ^2
   & 40-\frac{166}{3}\delta+\frac{173}{9}\delta ^2 \\
  \end{array}
  \right)$
  \end{widetext}
\end{itemize}

\subsection{$q^4 s^5:\{1234\}\{56789\}$}
\begin{itemize}
  \item $I=2, S=\frac{3}{2}(F=64)$:\\
  $H_{CS}=56-\frac{208}{3}\delta+32\delta^2$\\

  \item $I=2, S=\frac{1}{2}(F=\overline{35})$:\\
  $H_{CS}=40-\frac{160}{3}\delta+32\delta^2$\\

  \item $I=1, S=\frac{5}{2}(F=27)$:\\
  $H_{CS}=\frac{104}{3}-\frac{176}{3}\delta+32\delta^2$\\

  \item $I=1, S=\frac{3}{2}(F=27,64)$:\\
  $H_{CS}=\left(
  \begin{array}{cc}
  28-\frac{1172}{21}\delta+32 \delta ^2 & \frac{16 \sqrt{10}}{21}\delta \\
  \frac{16 \sqrt{10}}{21}\delta & 56-\frac{1712}{21}\delta+32 \delta ^2 \\
  \end{array}
  \right)$\\

  \item $I=1, S=\frac{1}{2}(F=27,35)$:\\
  $H_{CS}=\left(
  \begin{array}{cc}
  24-\frac{164}{3}\delta+32 \delta ^2 & -\frac{4}{3} \delta \\
  -\frac{4}{3}\delta & 40-\frac{212}{3}\delta+32 \delta ^2 \\
  \end{array}
  \right)$\\

  \item $I=0, S=\frac{3}{2}(F=10)$:\\
  $H_{CS}=20-\frac{172}{3}\delta+32\delta^2$\\

  \item $I=0, S=\frac{1}{2}(F=35)$:\\
  $H_{CS}=40-\frac{232}{3}\delta+32\delta^2$\\
\end{itemize}

\subsection{$q^3 s^6:\{123\}\{456789\}$}
\begin{itemize}
  \item $I=\frac{3}{2}, S=\frac{3}{2}(F=64)$:\\
  $H_{CS}=56-96\delta+48\delta^2$\\
  \item $I=\frac{1}{2}, S=\frac{1}{2}(F=35)$:\\
  $H_{CS}=40-96\delta+48\delta^2$\\
\end{itemize}

\section{Diquark configuration of a tribaryon}
\label{Diquark configuration of a tribaryon}

In this section, we analyze the diquark structure of a tribaryon in the flavor SU(3) broken case. By analyzing the diquark structure of a multiparticle system, we can reconstruct the diagonal elements of the color-spin or flavor-spin interaction by adding the contribution of the  corresponding interaction strength of all diquark configurations~\cite{Park:2019bsz}. Additionally, the information of the detailed diquark configurations would also be useful to analyze the interaction originating from the meson exchange between quarks~\cite{Glozman:1995fu}.

The way to calculate the diquark configuration is as follows. Let's examine a  diquark component between $u$,$d$ quarks. Considering the symmetry, it is sufficient to calculate the diquark component between 1 and 2. First, for a corresponding flavor state, assemble the Young-Yamanouchi basis into two sets  that are symmetric and antisymmetric under the exchange between 1 and 2, respectively. Similarly, we can also collect the symmetric and antisymmetric basis set for color and spin states. Then, by taking the sum of the expectation value, we can calculate the diquark configuration for $u$,$d$ quarks as follows.
\begin{align}
  P_{12}(F,I,S)=\sum_{F,C,S}\langle \Psi_{FCS}|{\cal \hat{P}}_{12}| \Psi_{FCS} \rangle
\label{diquark-equation}
\end{align}
with a projection operator,
${\cal \hat{P}}_{ij} \equiv |\psi^{\rm d}_{ij}  \rangle \langle \psi^{\rm d}_{ij} | \otimes \mathbf{1}$,
where $\psi^{\rm d}_{ij} $ is  the  wavefunction of the relevant diquark that satisfies a certain symmetry property, $\mathbf{1}$ is a unit operator acting on the particles other than $i$ and $j$, and $\Psi_{FCS}$ is the wave function of a tribaryon.

For the diquark structure involving $s$ quarks, let's consider $q^6 s^3$ case again. In this case, because of the symmetric property, we only need to consider $P_{67}(F,I,S)$ and $P_{89}(F,I,S)$ for the diquark structure involving $u$ and $s$ quarks, and $s$ quarks only, respectively. However, unless the particle label is 1 and 2, it is not easy to find the symmetric or antisymmetric basis set by just looking at the Young-Yamanouchi basis. Still, there is a useful way to find the symmetric or antisymmetric basis set. The eigenvalues of the elements $(i,j)$ of the irreducible representation of the corresponding Young diagram are 1 or -1. Then, the eigenvectors with eigenvalue of 1 become symmetric basis set and the opposite case becomes the antisymmetric basis set. $\psi_{ij}^d$ in Eq.(\ref{diquark-equation}) is an element of these basis sets. Now, using the Eq.(\ref{diquark-equation}), we can calculate the corresponding diquark component with a certain symmetric property. Additionally, the number of symmetric or antisymmetric basis between any $i$ and $j$ in a specific Young diagram is the same as the corresponding number between 1 and 2. The numbers of symmetric and antisymmetric basis for $S_9$ Young diagrams are written in Table \ref{symmetric-antisymmetric}.

\begin{widetext}
\begin{center}
\begin{table}
  \begin{tabular}{|c|c|c|c|c|c|c|c|c|c|c|c|c|c|c|c|c|}
  \hline
  & [9] & [81] & [72] & [711] & [63] & [621] & [6111] & [54] & [531] & [522] & [5211] & [51111] & [441] & [432] & [4311] & [333] \\
  \hline
  Symmetric & 1 & 7 & 21 & 21 & 34 & 70 & 35 & 28 & 99 & 70 & 105 & 35 & 49 & 91 & 111 & 21 \\
  \hline
  Antisymmetric & 0 & 1 & 6 & 7 & 14 & 35 & 21 & 14 & 63 & 50 & 84 & 35 & 35 & 77 & 105 & 21 \\
  \hline
  \end{tabular}
  \caption{The number of basis for symmetric and antisymmetric representation. For conjugate Young diagram, the numbers of symmetric and antisymmetric basis are reversed.}
  \label{symmetric-antisymmetric}
\end{table}
\end{center}
\end{widetext}

\begin{table}
\begin{center}
\begin{tabular}{c|c|c|c|c}
\hline
\hline
& \multicolumn{4}{c}{$q_i q_j$} \\
\hline
Flavor & $A(\bar{3})$ & $S(6)$ & $A(\bar{3})$ &$S(6)$ \\
\hline
Color & $A(\bar{3})$ & $A(\bar{3})$  & $S(6)$  & $S(6)$ \\
\hline
Spin & $A(1)$ & $S(3)$ & $S(3)$ & $A(1)$ \\
\hline
$- \lambda_i \lambda_j \sigma_i \cdot \sigma_j$ & $-8$& $\frac{8}{3}$ &  -$\frac{4}{3}$ & $4$ \\
\hline
\end{tabular}
\end{center}
\caption{The classification of two quark interaction. The two quark state is determined to satisfy Pauli exclusion principle. We denote the antisymmetric and symmetric state as $A$ and $S$ respectively. The symbols inside the parenthesis represents the  multiplet state.}
\label{two-quark-interaction}
\end{table}

Table \ref{two-quark-interaction} shows the basic building blocks composed of two quarks, henceforth called diquarks, and their corresponding strengths for the color-spin interaction.  Treating the quarks as identical particles, the diquarks can be in one of the four possible configurations  depending on the symmetry property under the exchange of the flavor, color and spin states.

\begin{center}
\begin{table}
  \begin{tabular}{|c|c|c|c|c|}
    \hline
    $q^9$ & \multicolumn{4}{|c|}{$i,j=1$-9} \\
    \hline  \hline
    Flavor & $A$ & $S$ & $A$ & $S$   \\
    \hline
    Color  & $A$ & $A$ & $S$ & $S$   \\
    \hline
    Spin  & $A$ & $S$ & $S$ & $A$   \\
    \hline
    $P_{ij}(F$=1,$S$=$\frac{3}{2}$) & $\frac{7}{48}$ & $\frac{17}{48}$ & $\frac{17}{48}$ & $\frac{7}{48}$ \\
    \hline
    $P_{ij}(F$=1,$S$=$\frac{5}{2}$) & $\frac{1}{9}$ & $\frac{7}{18}$ & $\frac{7}{18}$ & $\frac{1}{9}$ \\
    \hline
    $P_{ij}(F$=1,$S$=$\frac{9}{2}$) & 0 & $\frac{1}{2}$ & $\frac{1}{2}$ & 0 \\
    \hline
    $P_{ij}(F$=8,$S$=$\frac{1}{2}$) & $\frac{7}{48}$ & $\frac{17}{48}$ & $\frac{5}{16}$ & $\frac{3}{16}$ \\
    \hline
    $P_{ij}(F$=8,$S$=$\frac{3}{2}$) & $\frac{1}{8}$ & $\frac{3}{8}$ & $\frac{1}{3}$ & $\frac{1}{6}$ \\
    \hline
    $P_{ij}(F$=8,$S$=$\frac{5}{2}$) & $\frac{13}{144}$ & $\frac{59}{144}$ & $\frac{53}{144}$ & $\frac{19}{144}$ \\
    \hline
    $P_{ij}(F$=8,$S$=$\frac{7}{2}$) & $\frac{1}{24}$ & $\frac{11}{24}$ & $\frac{5}{12}$ & $\frac{1}{12}$ \\
    \hline
    $P_{ij}(F$=10,$S$=$\frac{3}{2}$) & $\frac{5}{48}$ & $\frac{19}{48}$ & $\frac{5}{16}$ & $\frac{3}{16}$ \\
    \hline
    $P_{ij}(F$=$\overline{10}$,$S$=$\frac{3}{2}$) & $\frac{5}{48}$ & $\frac{19}{48}$ & $\frac{5}{16}$ & $\frac{3}{16}$ \\
    \hline
    $P_{ij}(F$=27,$S$=$\frac{1}{2}$) & $\frac{1}{9}$ & $\frac{7}{18}$ & $\frac{5}{18}$ & $\frac{2}{9}$ \\
    \hline
    $P_{ij}(F$=27,$S$=$\frac{3}{2}$) & $\frac{13}{144}$ & $\frac{59}{144}$ & $\frac{43}{144}$ & $\frac{29}{144}$ \\
    \hline
    $P_{ij}(F$=27,$S$=$\frac{5}{2}$) & $\frac{1}{18}$ & $\frac{4}{9}$ & $\frac{1}{3}$ & $\frac{1}{6}$ \\
    \hline
    $P_{ij}(F$=35,$S$=$\frac{1}{2}$) & $\frac{1}{12}$ & $\frac{5}{12}$ & $\frac{1}{4}$ & $\frac{1}{4}$ \\
    \hline
    $P_{ij}(F$=$\overline{35}$,$S$=$\frac{1}{2}$) & $\frac{1}{12}$ & $\frac{5}{12}$ & $\frac{1}{4}$ & $\frac{1}{4}$ \\
    \hline
    $P_{ij}(F$=64,$S$=$\frac{3}{2}$) & $\frac{1}{24}$ & $\frac{11}{24}$ & $\frac{1}{4}$ & $\frac{1}{4}$ \\
    \hline
  \end{tabular}
  \caption{The probability $P_{ij}$ of a tribaryon for $(i,j)$ diquark pairs with  $S$,$A$ and $M$ representing symmetric ($S$),
   antisymmetric ($AS$)  and mixed ($M$) combinations in the  flavor SU(3) symmetric case.}
  \label{probability-diquark-symmetric}
\end{table}
\end{center}

\begin{center}
\begin{table}
  \begin{tabular}{|c|c|c|c|c|}
    \hline
    $q^9$ & \multicolumn{4}{|c|}{$i,j=1$-9} \\
    \hline  \hline
    Flavor & $A$ & $S$ & $A$ & $S$   \\
    \hline
    Color  & $A$ & $A$ & $S$ & $S$   \\
    \hline
    Spin  & $A$ & $S$ & $S$ & $A$   \\
    \hline
    $P_{ij}(F$=$\overline{35}$,$I=\frac{1}{2}$,$S$=$\frac{1}{2}$) & $\frac{1}{12}$ & $\frac{5}{12}$ & $\frac{1}{4}$ & $\frac{1}{4}$ \\
    \hline
    $P_{ij}(F$=64,$I=\frac{3}{2}$,$S$=$\frac{3}{2}$) & $\frac{1}{24}$ & $\frac{11}{24}$ & $\frac{1}{4}$ & $\frac{1}{4}$ \\
    \hline
  \end{tabular}
  \caption{The probability $P_{ij}$ of $q^9$ for $(i,j)$ diquark pairs with  $S$,$A$ and $M$ representing symmetric ($S$),
   antisymmetric ($AS$)  and mixed ($M$) combinations in the  flavor SU(3) broken case.}
  \label{probability-diquark-0}
\end{table}
\end{center}

We represent the diquark configuration of a tribaryon for all possible states in Table \ref{probability-diquark-symmetric}-\ref{probability-diquark-6}. Table \ref{probability-diquark-symmetric} shows the result in the flavor SU(3) symmetric case. Additionally, it should be noted that $u$ and $s$ are treated as different flavors and hence do  not have to satisfy the Pauli exclusion principle in the flavor SU(3) broken case. So there are in total 8 cases as shown in different columns to represent all possible diquark structure involving $u$ and $s$ quarks. By comparing each tables with Table \ref{probability-diquark-symmetric}, we can analyze how the diquark components change when the flavor symmetry is broken.

The most attractive tribaryon state is related to configurations where there are large fraction of attractive diquark configurations.
As an example, the way to find the most attractive channel is to identify  configuration that has large contributions from the A-color(antisymmetric) and A-spin diquark state. For the flavor SU(3) symmetric case, as we can see in Table \ref{probability-diquark-symmetric}, the  flavor singlet with $S=\frac{3}{2}$ and flavor octet with $S=\frac{1}{2}$ have the largest A-color and A-spin  diquark state contribution. We can check that the flavor singlet state is the most attractive state as can be seen by letting $\delta = 0$ in the color-spin matrix elements given in Sec.\ref{Color-spin interaction matrix elements} for the flavor symmetric case. For flavor singlet with $S=\frac{3}{2}$, $H_{CS}$ is -4 and for flavor octet with $S=\frac{1}{2}$, $H_{CS}$ is 4. This difference comes from the remaining diquark components other than A-color and A-spin states. As can be seen in Table \ref{two-quark-interaction}, there are two attractive diquarks. The reason why flavor singlet state is more attractive is that compared to the flavor octet case, it has larger contributions from the other remaining, although with less degree, attractive diquark components, which is the S-color(symmetric) and S-spin state. Therefore, the flavor states which have large components of A-color, A-spin and S-color, S-spin states lead to attractive configurations. For the flavor symmetry broken case, it is still useful to search for  the diquark components between $u$,$d$ quarks to find the most attractive channel. The followings are the most attractive diagonal channels for all possible strangeness  quantum numbers, up to 6 strange quarks, in the flavor SU(3) broken case.

\begin{itemize}
  \item $q^9$ : $(F,I,S)=(\overline{35},\frac{1}{2},\frac{1}{2})$
  \item $q^8 s$ : $(F,I,S)=(\overline{10},0,\frac{3}{2})$, $(\overline{35},0,\frac{1}{2})$
  \item $q^7 s^2$ : $(F,I,S)=(27,\frac{1}{2},\frac{1}{2})$
  \item $q^6 s^3$ : $(F,I,S)=(27,0,\frac{1}{2})$
  \item $q^5 s^4$ : $(F,I,S)=(8,\frac{1}{2},\frac{1}{2})$
  \item $q^4 s^5$ : $(F,I,S)=(10,0,\frac{3}{2}),(35,0,\frac{1}{2})$
  \item $q^3 s^6$ : $(F,I,S)=(35,\frac{1}{2},\frac{1}{2})$
\end{itemize}

To obtained the lowest eigenvalue for the present case, we need to consider the mixing effects between different SU(3) flavor channels, and diagonalize the whole matrix components in Sec.\ref{Color-spin interaction matrix elements}.

\begin{widetext}
\begin{center}
\begin{table}
  \begin{tabular}{|c|c|c|c|c|}
    \hline
    $q^8 s$ & \multicolumn{4}{|c|}{$i,j=1$-8} \\
    \hline  \hline
    Flavor & $A$ & $S$ & $A$ & $S$   \\
    \hline
    Color  & $A$ & $A$ & $S$ & $S$   \\
    \hline
    Spin  & $A$ & $S$ & $S$ & $A$   \\
    \hline
    $P_{ij}(F$=64,$I$=2,$S$=$\frac{3}{2}$) & $\frac{1}{28}$ & $\frac{13}{28}$ & $\frac{3}{14}$ & $\frac{2}{7}$ \\
    \hline
    $P_{ij}(F$=35,$I$=2,$S$=$\frac{1}{2}$) & $\frac{1}{28}$ & $\frac{13}{28}$ & $\frac{3}{14}$ & $\frac{2}{7}$ \\
    \hline
    $P_{ij}(F$=27,$I$=1,$S$=$\frac{5}{2}$) & $\frac{1}{28}$ & $\frac{13}{28}$ & $\frac{2}{7}$ & $\frac{3}{14}$ \\
    \hline
    $P_{ij}(F$=64,$I$=1,$S$=$\frac{3}{2}$) & $\frac{9}{196}$ & $\frac{89}{196}$ & $\frac{27}{98}$ & $\frac{11}{49}$ \\
    \hline
    $P_{ij}(F$=27,$I$=1,$S$=$\frac{3}{2}$) & $\frac{3}{49}$ & $\frac{43}{98}$ & $\frac{51}{196}$ & $\frac{47}{196}$ \\
    \hline
    $P_{ij}(F$=$\overline{35}$,$I$=1,$S$=$\frac{1}{2}$) & $\frac{9}{112}$ & $\frac{47}{112}$ & $\frac{27}{112}$ & $\frac{29}{112}$ \\
    \hline
    $P_{ij}(F$=27,$I$=1,$S$=$\frac{1}{2}$) & $\frac{9}{112}$ & $\frac{47}{112}$ & $\frac{27}{112}$ & $\frac{29}{112}$ \\
    \hline
    $P_{ij}(F$=$\overline{10}$,$I$=0,$S$=$\frac{3}{2}$) & $\frac{5}{56}$ & $\frac{23}{56}$ & $\frac{15}{56}$ & $\frac{13}{56}$ \\
    \hline
    $P_{ij}(F$=$\overline{35}$,$I$=0,$S$=$\frac{1}{2}$) & $\frac{5}{56}$ & $\frac{23}{56}$ & $\frac{15}{56}$ & $\frac{13}{56}$ \\
    \hline
  \end{tabular}
  \begin{tabular}{|c|c|c|c|c|c|c|c|}
    \hline
    \multicolumn{8}{|c|}{$i=1$-8,$j=9$} \\
    \hline  \hline
    $A$ & $S$ & $A$ & $S$  & $A$ & $S$ & $A$ & $S$ \\
    \hline
    $A$ & $A$ & $S$ & $S$ & $S$ & $S$ & $A$ & $A$  \\
    \hline
    $A$ & $S$ & $S$ & $A$ & $A$ & $S$ & $S$ & $A$  \\
    \hline
    $\frac{5}{128}$ & $\frac{45}{128}$ & $\frac{15}{64}$ & $\frac{3}{64}$ & $\frac{5}{64}$ & $\frac{9}{64}$ & $\frac{11}{128}$ & $\frac{3}{128}$ \\
    \hline
    $\frac{41}{224}$ & $\frac{3}{32}$ & $\frac{3}{14}$ & $\frac{3}{56}$ & $\frac{1}{14}$ & $\frac{9}{56}$ & $\frac{5}{32}$ & $\frac{15}{224}$ \\
    \hline
    $\frac{7}{96}$ & $\frac{41}{288}$ & $\frac{23}{72}$ & 0 & 0 & $\frac{13}{72}$ & $\frac{67}{288}$ & $\frac{5}{96}$ \\
    \hline
    $\frac{9}{896}$ & $\frac{369}{896}$ & $\frac{27}{448}$ & $\frac{127}{448}$ & $\frac{25}{448}$ & $\frac{45}{448}$ & $\frac{55}{896}$ & $\frac{15}{896}$ \\
    \hline
    $\frac{983}{8064}$ & $\frac{107}{896}$ & $\frac{727}{2688}$ & $\frac{185}{8064}$ & $\frac{355}{8064}$ & $\frac{437}{2688}$ & $\frac{169}{896}$ & $\frac{565}{8064}$ \\
    \hline
    $\frac{9}{128}$ & $\frac{45}{128}$ & $\frac{27}{128}$ & $\frac{23}{128}$ & $\frac{5}{128}$ & $\frac{9}{128}$ & $\frac{7}{128}$ & $\frac{3}{128}$ \\
    \hline
    $\frac{167}{1152}$ & $\frac{41}{384}$ & $\frac{95}{384}$ & $\frac{41}{1152}$ & $\frac{67}{1152}$ & $\frac{61}{384}$ & $\frac{67}{384}$ & $\frac{85}{1152}$ \\
    \hline
    $\frac{25}{256}$ & $\frac{33}{256}$ & $\frac{75}{256}$ & $\frac{3}{256}$ & $\frac{5}{256}$ & $\frac{45}{256}$ & $\frac{55}{256}$ & $\frac{15}{256}$ \\
    \hline
    $\frac{1}{64}$ & $\frac{21}{64}$ & $\frac{3}{64}$ & $\frac{15}{64}$ & $\frac{5}{64}$ & $\frac{9}{64}$ & $\frac{7}{64}$ & $\frac{3}{64}$ \\
    \hline
  \end{tabular}
  \caption{Same as Table \ref{probability-diquark-0}. for $q^8s$}
  \label{probability-diquark-1}
\end{table}
\end{center}

\begin{center}
\begin{table}
  \begin{tabular}{|c|c|c|c|c|}
    \hline
    $q^7 s^2$ & \multicolumn{4}{|c|}{$i,j=1$-7} \\
    \hline  \hline
    Flavor & $A$ & $S$ & $A$ & $S$   \\
    \hline
    Color  & $A$ & $A$ & $S$ & $S$   \\
    \hline
    Spin  & $A$ & $S$ & $S$ & $A$   \\
    \hline
    $P_{ij}(F$=64,$I$=$\frac{5}{2}$,$S$=$\frac{3}{2}$) & $\frac{1}{42}$ & $\frac{1}{2}$ & $\frac{1}{7}$ & $\frac{1}{3}$ \\
    \hline
    $P_{ij}(F$=35,$I$=$\frac{5}{2}$,$S$=$\frac{1}{2}$) & $\frac{1}{42}$ & $\frac{1}{2}$ & $\frac{1}{7}$ & $\frac{1}{3}$ \\
    \hline
    $P_{ij}(F$=27,$I$=$\frac{3}{2}$,$S$=$\frac{5}{2}$) & $\frac{1}{21}$ & $\frac{10}{21}$ & $\frac{5}{21}$ & $\frac{5}{21}$ \\
    \hline
    $P_{ij}(F$=10,$I$=$\frac{3}{2}$,$S$=$\frac{3}{2}$) & $\frac{3}{56}$ & $\frac{233}{504}$ & $\frac{13}{56}$ & $\frac{127}{504}$ \\
    \hline
    $P_{ij}(F$=27,$I$=$\frac{3}{2}$,$S$=$\frac{3}{2}$) & $\frac{71}{1176}$ & $\frac{25}{56}$ & $\frac{265}{1176}$ & $\frac{15}{56}$ \\
    \hline
    $P_{ij}(F$=64,$I$=$\frac{3}{2}$,$S$=$\frac{3}{2}$) & $\frac{2}{49}$ & $\frac{29}{63}$ & $\frac{12}{49}$ & $\frac{16}{63}$ \\
    \hline
    $P_{ij}(F$=27,$I$=$\frac{3}{2}$,$S$=$\frac{1}{2}$) & $\frac{11}{168}$ & $\frac{25}{56}$ & $\frac{37}{168}$ & $\frac{15}{56}$ \\
    \hline
    $P_{ij}(F$=35,$I$=$\frac{3}{2}$,$S$=$\frac{1}{2}$) & $\frac{3}{56}$ & $\frac{227}{504}$ & $\frac{13}{56}$ & $\frac{19}{72}$ \\
    \hline
    $P_{ij}(F$=$\overline{35}$,$I$=$\frac{3}{2}$,$S$=$\frac{1}{2}$) & $\frac{1}{14}$ & $\frac{55}{126}$ & $\frac{3}{14}$ & $\frac{5}{18}$ \\
    \hline
    $P_{ij}(F$=8,$I$=$\frac{1}{2}$,$S$=$\frac{7}{2}$) & $\frac{1}{42}$ & $\frac{1}{2}$ & $\frac{1}{3}$ & $\frac{1}{7}$ \\
    \hline
    $P_{ij}(F$=27,$I$=$\frac{1}{2}$,$S$=$\frac{5}{2}$) & $\frac{1}{30}$ & $\frac{103}{210}$ & $\frac{34}{105}$ & $\frac{16}{105}$ \\
    \hline
    $P_{ij}(F$=8,$I$=$\frac{1}{2}$,$S$=$\frac{5}{2}$) & $\frac{31}{420}$ & $\frac{9}{20}$ & $\frac{17}{60}$ & $\frac{27}{140}$ \\
    \hline
    $P_{ij}(F$=8,$I$=$\frac{1}{2}$,$S$=$\frac{3}{2}$) & $\frac{31}{420}$ & $\frac{179}{420}$ & $\frac{17}{60}$ & $\frac{13}{60}$ \\
    \hline
    $P_{ij}(F$=$\overline{10}$,$I$=$\frac{1}{2}$,$S$=$\frac{3}{2}$) & $\frac{5}{56}$ & $\frac{31}{72}$ & $\frac{15}{56}$ & $\frac{107}{504}$ \\
    \hline
    $P_{ij}(F$=27,$I$=$\frac{1}{2}$,$S$=$\frac{3}{2}$) & $\frac{421}{5880}$ & $\frac{883}{1960}$ & $\frac{1679}{5880}$ & $\frac{377}{1960}$ \\
    \hline
    $P_{ij}(F$=64,$I$=$\frac{1}{2}$,$S$=$\frac{3}{2}$) & $\frac{5}{98}$ & $\frac{401}{882}$ & $\frac{15}{49}$ & $\frac{83}{441}$ \\
    \hline
    $P_{ij}(F$=8,$I$=$\frac{1}{2}$,$S$=$\frac{1}{2}$) & $\frac{3}{28}$ & $\frac{101}{252}$ & $\frac{1}{4}$ & $\frac{61}{252}$ \\
    \hline
    $P_{ij}(F$=27,$I$=$\frac{1}{2}$,$S$=$\frac{1}{2}$) & $\frac{17}{168}$ & $\frac{23}{56}$ & $\frac{43}{168}$ & $\frac{13}{56}$ \\
    \hline
    $P_{ij}(F$=$\overline{35}$,$I$=$\frac{1}{2}$,$S$=$\frac{1}{2}$) & $\frac{5}{56}$ & $\frac{209}{504}$ & $\frac{15}{56}$ & $\frac{115}{504}$ \\
    \hline
  \end{tabular}
  \begin{tabular}{|c|c|}
    \hline
    \multicolumn{2}{|c|}{$i,j$=8,9} \\
    \hline  \hline
     ~$S$~ & $S$   \\
    \hline
     $S$ & $A$   \\
    \hline
     $A$ & $S$   \\
    \hline
    1 & 0  \\
    \hline
    1 & 0  \\
    \hline
    1 & 0  \\
    \hline
    $\frac{5}{6}$ & $\frac{1}{6}$  \\
    \hline
    $\frac{9}{14}$ & $\frac{5}{14}$  \\
    \hline
    $\frac{11}{21}$ & $\frac{10}{21}$  \\
    \hline
    $\frac{3}{4}$ & $\frac{1}{4}$  \\
    \hline
    $\frac{7}{12}$ & $\frac{5}{12}$  \\
    \hline
    $\frac{2}{3}$ & $\frac{1}{3}$  \\
    \hline
    1 & 0  \\
    \hline
    1 & 0  \\
    \hline
    1 & 0  \\
    \hline
    $\frac{1}{2}$ & $\frac{1}{2}$  \\
    \hline
    $\frac{11}{12}$ & $\frac{1}{12}$  \\
    \hline
    $\frac{27}{28}$ & $\frac{1}{28}$  \\
    \hline
    $\frac{13}{21}$ & $\frac{8}{21}$  \\
    \hline
    $\frac{2}{3}$ & $\frac{1}{3}$  \\
    \hline
    $\frac{3}{4}$ & $\frac{1}{4}$  \\
    \hline
    $\frac{7}{12}$ & $\frac{5}{12}$  \\
    \hline
  \end{tabular}
  \begin{tabular}{|c|c|c|c|c|c|c|c|}
    \hline
    \multicolumn{8}{|c|}{$i=1$-7,$j=8,9$} \\
    \hline  \hline
    $A$ & $S$ & $A$ & $S$  & $A$ & $S$ & $A$ & $S$ \\
    \hline
    $A$ & $A$ & $S$ & $S$ & $S$ & $S$ & $A$ & $A$  \\
    \hline
    $A$ & $S$ & $S$ & $A$ & $A$ & $S$ & $S$ & $A$  \\
    \hline
    $\frac{3}{70}$ & $\frac{17}{70}$ & $\frac{9}{35}$ & $\frac{2}{35}$ & $\frac{3}{35}$ & $\frac{6}{35}$ & $\frac{4}{35}$ & $\frac{1}{35}$ \\
    \hline
    $\frac{489}{3920}$ & $\frac{11}{112}$ & $\frac{243}{980}$ & $\frac{59}{980}$ & $\frac{81}{980}$ & $\frac{177}{980}$ & $\frac{17}{112}$ & $\frac{211}{3920}$ \\
    \hline
    $\frac{11}{252}$ & $\frac{10}{63}$ & $\frac{2}{7}$ & $\frac{1}{36}$ & $\frac{11}{252}$ & $\frac{3}{14}$ & $\frac{25}{126}$ & $\frac{1}{36}$ \\
    \hline
    $\frac{59}{448}$ & $\frac{395}{4032}$ & $\frac{129}{448}$ & $\frac{145}{4032}$ & $\frac{227}{4032}$ & $\frac{75}{448}$ & $\frac{673}{4032}$ & $\frac{25}{448}$ \\
    \hline
    $\frac{1121}{14112}$ & $\frac{305}{2016}$ & $\frac{3527}{14112}$ & $\frac{71}{2016}$ & $\frac{781}{14112}$ & $\frac{2539}{14112}$ & $\frac{2635}{14112}$ & $\frac{877}{14112}$ \\
    \hline
    $\frac{67}{3920}$ & $\frac{1621}{5040}$ & $\frac{201}{1960}$ & $\frac{391}{2520}$ & $\frac{1283}{17640}$ & $\frac{339}{1960}$ & $\frac{4553}{35280}$ & $\frac{113}{3920}$ \\
    \hline
    $\frac{103}{1008}$ & $\frac{121}{1008}$ & $\frac{229}{1008}$ & $\frac{67}{1008}$ & $\frac{43}{504}$ & $\frac{79}{504}$ & $\frac{79}{504}$ & $\frac{43}{504}$ \\
    \hline
    $\frac{43}{560}$ & $\frac{53}{252}$ & $\frac{69}{560}$ & $\frac{167}{1260}$ & $\frac{61}{720}$ & $\frac{6}{35}$ & $\frac{145}{1008}$ & $\frac{2}{35}$ \\
    \hline
    $\frac{1}{14}$ & $\frac{71}{252}$ & $\frac{3}{14}$ & $\frac{37}{252}$ & $\frac{1}{18}$ & $\frac{3}{28}$ & $\frac{11}{126}$ & $\frac{1}{28}$ \\
    \hline
    $\frac{3}{70}$ & $\frac{9}{70}$ & $\frac{13}{35}$ & 0 & 0 & $\frac{1}{5}$ & $\frac{8}{35}$ & $\frac{1}{35}$ \\
    \hline
    $\frac{199}{5040}$ & $\frac{4651}{25200}$ & $\frac{11}{60}$ & $\frac{691}{6300}$ & $\frac{569}{6300}$ & $\frac{79}{420}$ & $\frac{3809}{25200}$ & $\frac{269}{5040}$ \\
    \hline
    $\frac{9}{112}$ & $\frac{311}{2800}$ & $\frac{187}{560}$ & $\frac{7}{400}$ & $\frac{13}{400}$ & $\frac{3}{16}$ & $\frac{5499}{2800}$ & $\frac{23}{560}$ \\
    \hline
    $\frac{15}{112}$ & $\frac{883}{8400}$ & $\frac{157}{560}$ & $\frac{197}{8400}$ & $\frac{373}{8400}$ & $\frac{17}{112}$ & $\frac{221}{1200}$ & $\frac{43}{560}$ \\
    \hline
    $\frac{5}{64}$ & $\frac{611}{4032}$ & $\frac{15}{64}$ & $\frac{361}{4032}$ & $\frac{275}{4032}$ & $\frac{75}{448}$ & $\frac{625}{4032}$ & $\frac{25}{448}$ \\
    \hline
    $\frac{8507}{141120}$ & $\frac{122503}{705600}$ & $\frac{22937}{141120}$ & $\frac{85117}{705600}$ & $\frac{74903}{705600}$ & $\frac{24979}{141120}$ & $\frac{95477}{705600}$ & $\frac{9097}{141120}$ \\
    \hline
    $\frac{1}{98}$ & $\frac{331}{882}$ & $\frac{3}{49}$ & $\frac{118}{441}$ & $\frac{29}{441}$ & $\frac{6}{49}$ & $\frac{34}{441}$ & $\frac{1}{49}$ \\
    \hline
    $\frac{79}{560}$ & $\frac{481}{5040}$ & $\frac{153}{560}$ & $\frac{167}{5040}$ & $\frac{313}{5040}$ & $\frac{87}{560}$ & $\frac{839}{5040}$ & $\frac{41}{560}$ \\
    \hline
    $\frac{1409}{20160}$ & $\frac{3641}{20160}$ & $\frac{3083}{20160}$ & $\frac{2291}{20160}$ & $\frac{1849}{20160}$ & $\frac{511}{2880}$ & $\frac{3019}{20160}$ & $\frac{1291}{20160}$ \\
    \hline
    $\frac{23}{448}$ & $\frac{1367}{4032}$ & $\frac{69}{448}$ & $\frac{901}{4032}$ & $\frac{191}{4032}$ & $\frac{39}{448}$ & $\frac{277}{4032}$ & $\frac{13}{448}$ \\
    \hline
  \end{tabular}
  \caption{Same as Table \ref{probability-diquark-0}. for $q^7s^2$}
  \label{probability-diquark-2}
\end{table}
\end{center}

\begin{center}
\begin{table}
  \begin{tabular}{|c|c|c|c|c|}
    \hline
    $q^6 s^3$ & \multicolumn{4}{|c|}{$i,j=1$-6} \\
    \hline  \hline
    Flavor & $A$ & $S$ & $A$ & $S$   \\
    \hline
    Color  & $A$ & $A$ & $S$ & $S$   \\
    \hline
    Spin  & $A$ & $S$ & $S$ & $A$   \\
    \hline
    $P_{ij}(F$=64,$I$=3,$S$=$\frac{3}{2}$) & 0 & $\frac{3}{5}$ & 0 & $\frac{2}{5}$ \\
    \hline
    $P_{ij}(F$=27,$I$=2,$S$=$\frac{5}{2}$) & $\frac{1}{15}$ & $\frac{8}{15}$ & $\frac{2}{15}$ & $\frac{4}{15}$ \\
    \hline
    $P_{ij}(F$=27,$I$=2,$S$=$\frac{3}{2}$) & $\frac{23}{420}$ & $\frac{73}{140}$ & $\frac{61}{420}$ & $\frac{39}{140}$ \\
    \hline
    $P_{ij}(F$=64,$I$=2,$S$=$\frac{3}{2}$) & $\frac{1}{53}$ & $\frac{52}{105}$ & $\frac{6}{35}$ & $\frac{32}{105}$ \\
    \hline
    $P_{ij}(F$=27,$I$=2,$S$=$\frac{1}{2}$) & $\frac{1}{30}$ & $\frac{1}{2}$ & $\frac{1}{6}$ & $\frac{3}{10}$ \\
    \hline
    $P_{ij}(F$=35,$I$=2,$S$=$\frac{1}{2}$) & $\frac{1}{20}$ & $\frac{29}{60}$ & $\frac{3}{20}$ & $\frac{19}{60}$ \\
    \hline
    $P_{ij}(F$=$\overline{35}$,$I$=2,$S$=$\frac{1}{2}$) & $\frac{1}{20}$ & $\frac{29}{60}$ & $\frac{3}{20}$ & $\frac{19}{60}$ \\
    \hline
    $P_{ij}(F$=8,$I$=1,$S$=$\frac{7}{2}$) & $\frac{1}{15}$ & $\frac{8}{15}$ & $\frac{4}{15}$ & $\frac{2}{15}$ \\
    \hline
    $P_{ij}(F$=8,$I$=1,$S$=$\frac{5}{2}$) & $\frac{13}{300}$ & $\frac{51}{100}$ & $\frac{29}{100}$ & $\frac{47}{300}$ \\
    \hline
    $P_{ij}(F$=27,$I$=1,$S$=$\frac{5}{2}$) & $\frac{1}{25}$ & $\frac{38}{75}$ & $\frac{22}{75}$ & $\frac{4}{25}$ \\
    \hline
    $P_{ij}(F$=8,$I$=1,$S$=$\frac{3}{2}$) & $\frac{19}{150}$ & $\frac{197}{450}$ & $\frac{31}{150}$ & $\frac{103}{450}$ \\
    \hline
    $P_{ij}(F$=10,$I$=1,$S$=$\frac{3}{2}$) & $\frac{1}{12}$ & $\frac{25}{54}$ & $\frac{1}{4}$ & $\frac{11}{54}$ \\
    \hline
    $P_{ij}(F$=$\overline{10}$,$I$=1,$S$=$\frac{3}{2}$) & $\frac{1}{12}$ & $\frac{25}{54}$ & $\frac{1}{4}$ & $\frac{11}{54}$ \\
    \hline
    $P_{ij}(F$=27,$I$=1,$S$=$\frac{3}{2}$) & $\frac{53}{700}$ & $\frac{949}{2100}$ & $\frac{541}{2100}$ & $\frac{451}{2100}$ \\
    \hline
    $P_{ij}(F$=64,$I$=1,$S$=$\frac{3}{2}$) & $\frac{1}{21}$ & $\frac{442}{945}$ & $\frac{2}{7}$ & $\frac{188}{945}$ \\
    \hline
    $P_{ij}(F$=8,$I$=1,$S$=$\frac{1}{2}$) & $\frac{1}{12}$ & $\frac{241}{540}$ & $\frac{1}{4}$ & $\frac{119}{540}$ \\
    \hline
    $P_{ij}(F$=27,$I$=1,$S$=$\frac{1}{2}$) & $\frac{1}{10}$ & $\frac{13}{30}$ & $\frac{7}{30}$ & $\frac{7}{30}$ \\
    \hline
    $P_{ij}(F$=35,$I$=1,$S$=$\frac{1}{2}$) & $\frac{1}{12}$ & $\frac{47}{108}$ & $\frac{1}{4}$ & $\frac{25}{108}$ \\
    \hline
    $P_{ij}(F$=$\overline{35}$,$I$=1,$S$=$\frac{1}{2}$) & $\frac{1}{12}$ & $\frac{47}{108}$ & $\frac{1}{4}$ & $\frac{25}{108}$ \\
    \hline
    $P_{ij}(F$=1,$I$=0,$S$=$\frac{9}{2}$) & 0 & $\frac{3}{5}$ & $\frac{2}{5}$ & 0 \\
    \hline
    $P_{ij}(F$=8,$I$=0,$S$=$\frac{7}{2}$) & 0 & $\frac{3}{5}$ & $\frac{2}{5}$ & 0 \\
    \hline
    $P_{ij}(F$=1,$I$=0,$S$=$\frac{5}{2}$) & $\frac{4}{45}$ & $\frac{7}{15}$ & $\frac{14}{45}$ & $\frac{2}{15}$ \\
    \hline
    $P_{ij}(F$=8,$I$=0,$S$=$\frac{5}{2}$) & $\frac{91}{900}$ & $\frac{29}{60}$ & $\frac{269}{900}$ & $\frac{7}{60}$ \\
    \hline
    $P_{ij}(F$=27,$I$=0,$S$=$\frac{5}{2}$) & $\frac{2}{75}$ & $\frac{8}{15}$ & $\frac{28}{75}$ & $\frac{1}{15}$ \\
    \hline
    $P_{ij}(F$=1,$I$=0,$S$=$\frac{3}{2}$) & $\frac{7}{60}$ & $\frac{17}{40}$ & $\frac{17}{60}$ & $\frac{7}{40}$ \\
    \hline
    $P_{ij}(F$=8,$I$=0,$S$=$\frac{3}{2}$) & $\frac{11}{150}$ & $\frac{1}{2}$ & $\frac{49}{150}$ & $\frac{1}{10}$ \\
    \hline
    $P_{ij}(F$=27,$I$=0,$S$=$\frac{3}{2}$) & $\frac{181}{2100}$ & $\frac{19}{40}$ & $\frac{659}{2100}$ & $\frac{1}{8}$ \\
    \hline
    $P_{ij}(F$=64,$I$=0,$S$=$\frac{3}{2}$) & $\frac{2}{35}$ & $\frac{7}{15}$ & $\frac{12}{35}$ & $\frac{2}{15}$ \\
    \hline
    $P_{ij}(F$=8,$I$=0,$S$=$\frac{1}{2}$) & $\frac{3}{20}$ & $\frac{5}{12}$ & $\frac{1}{4}$ & $\frac{11}{60}$ \\
    \hline
    $P_{ij}(F$=27,$I$=0,$S$=$\frac{1}{2}$) & $\frac{2}{15}$ & $\frac{2}{5}$ & $\frac{4}{15}$ & $\frac{1}{5}$ \\
    \hline
  \end{tabular}
  \begin{tabular}{|c|c|}
    \hline
    \multicolumn{2}{|c|}{$i,j$=7-9} \\
    \hline  \hline
     ~$S$~ & $S$   \\
    \hline
     $S$ & $A$   \\
    \hline
     $A$ & $S$   \\
    \hline
     1 & 0  \\
    \hline
     1 & 0  \\
    \hline
     $\frac{37}{42}$ & $\frac{5}{42}$  \\
    \hline
     $\frac{13}{21}$ & $\frac{8}{21}$  \\
    \hline
     $\frac{2}{3}$ & $\frac{1}{3}$  \\
    \hline
     $\frac{2}{3}$ & $\frac{1}{3}$  \\
    \hline
     $\frac{2}{3}$ & $\frac{1}{3}$  \\
    \hline
    1 & 0  \\
    \hline
     $\frac{11}{15}$ & $\frac{4}{15}$  \\
    \hline
     $\frac{11}{15}$ & $\frac{4}{15}$  \\
    \hline
     $\frac{37}{45}$ & $\frac{8}{45}$  \\
    \hline
     $\frac{79}{108}$ & $\frac{29}{108}$  \\
    \hline
     $\frac{79}{108}$ & $\frac{29}{108}$  \\
    \hline
     $\frac{67}{105}$ & $\frac{38}{105}$  \\
    \hline
     $\frac{109}{189}$ & $\frac{80}{189}$  \\
    \hline
     $\frac{35}{54}$ & $\frac{19}{54}$  \\
    \hline
     $\frac{2}{3}$ & $\frac{1}{3}$  \\
    \hline
     $\frac{16}{27}$ & $\frac{11}{27}$  \\
    \hline
     $\frac{16}{27}$ & $\frac{11}{27}$  \\
    \hline
     1 & 0  \\
    \hline
     1 & 0  \\
    \hline
     $\frac{7}{9}$ & $\frac{2}{9}$  \\
    \hline
     $\frac{83}{90}$ & $\frac{7}{90}$  \\
    \hline
     $\frac{4}{5}$ & $\frac{1}{5}$  \\
    \hline
     $\frac{17}{24}$ & $\frac{7}{24}$  \\
    \hline
    $\frac{13}{15}$ & $\frac{2}{15}$  \\
    \hline
    $\frac{677}{840}$ & $\frac{163}{840}$  \\
    \hline
    $\frac{13}{21}$ & $\frac{8}{21}$  \\
    \hline
    $\frac{5}{6}$ & $\frac{1}{6}$  \\
    \hline
    $\frac{2}{3}$ & $\frac{1}{3}$  \\
    \hline
  \end{tabular}
  \begin{tabular}{|c|c|c|c|c|c|c|c|}
    \hline
    \multicolumn{8}{|c|}{$i=1$-6,$j=7$-9} \\
    \hline  \hline
    $A$ & $S$ & $A$ & $S$  & $A$ & $S$ & $A$ & $S$ \\
    \hline
    $A$ & $A$ & $S$ & $S$ & $S$ & $S$ & $A$ & $A$  \\
    \hline
    $A$ & $S$ & $S$ & $A$ & $A$ & $S$ & $S$ & $A$  \\
    \hline
    $\frac{7}{144}$ & $\frac{5}{48}$ & $\frac{7}{24}$ & $\frac{5}{72}$ & $\frac{7}{72}$ & $\frac{5}{24}$ & $\frac{7}{48}$ & $\frac{5}{144}$ \\
    \hline
    $\frac{43}{1296}$ & $\frac{139}{1296}$ & $\frac{221}{648}$ & $\frac{29}{648}$ & $\frac{43}{648}$ & $\frac{139}{648}$ & $\frac{221}{1296}$ & $\frac{29}{1296}$ \\
    \hline
    $\frac{761}{9072}$ & $\frac{281}{3024}$ & $\frac{439}{1512}$ & $\frac{67}{1134}$ & $\frac{52}{567}$ & $\frac{281}{1512}$ & $\frac{439}{3024}$ & $\frac{463}{9072}$ \\
    \hline
    $\frac{29}{1008}$ & $\frac{29}{112}$ & $\frac{29}{168}$ & $\frac{55}{504}$ & $\frac{37}{504}$ & $\frac{31}{168}$ & $\frac{143}{1008}$ & $\frac{31}{1008}$ \\
    \hline
    $\frac{79}{648}$ & $\frac{7}{72}$ & $\frac{109}{432}$ & $\frac{71}{1296}$ & $\frac{109}{1296}$ & $\frac{71}{432}$ & $\frac{11}{72}$ & $\frac{47}{648}$ \\
    \hline
    $\frac{43}{576}$ & $\frac{281}{1728}$ & $\frac{35}{192}$ & $\frac{163}{1728}$ & $\frac{149}{1728}$ & $\frac{37}{192}$ & $\frac{271}{1728}$ & $\frac{29}{576}$ \\
    \hline
    $\frac{5}{64}$ & $\frac{13}{64}$ & $\frac{15}{64}$ & $\frac{7}{64}$ & $\frac{41}{576}$ & $\frac{9}{64}$ & $\frac{67}{5768}$ & $\frac{3}{64}$ \\
    \hline
    $\frac{37}{2160}$ & $\frac{79}{720}$ & $\frac{47}{120}$ & $\frac{23}{1080}$ & $\frac{37}{1080}$ & $\frac{79}{360}$ & $\frac{47}{240}$ & $\frac{23}{2160}$ \\
    \hline
    $\frac{251}{2700}$ & $\frac{173}{1800}$ & $\frac{379}{1200}$ & $\frac{377}{10800}$ & $\frac{643}{10800}$ & $\frac{643}{3600}$ & $\frac{307}{1800}$ & $\frac{139}{2700}$ \\
    \hline
    $\frac{421}{10800}$ & $\frac{1877}{10800}$ & $\frac{1147}{5400}$ & $\frac{419}{5400}$ & $\frac{421}{5400}$ & $\frac{1133}{5400}$ & $\frac{1843}{10800}$ & $\frac{419}{10800}$ \\
    \hline
    $\frac{251}{2700}$ & $\frac{121}{1350}$ & $\frac{379}{1200}$ & $\frac{149}{3600}$ & $\frac{773}{10800}$ & $\frac{643}{3600}$ & $\frac{107}{675}$ & $\frac{139}{2700}$ \\
    \hline
    $\frac{23}{270}$ & $\frac{209}{1620}$ & $\frac{7}{30}$ & $\frac{127}{1620}$ & $\frac{133}{1620}$ & $\frac{11}{60}$ & $\frac{251}{1620}$ & $\frac{29}{540}$ \\
    \hline
    $\frac{35}{432}$ & $\frac{173}{1296}$ & $\frac{35}{144}$ & $\frac{103}{1296}$ & $\frac{35}{432}$ & $\frac{25}{144}$ & $\frac{65}{432}$ & $\frac{25}{432}$ \\
    \hline
    $\frac{4397}{75600}$ & $\frac{12829}{75600}$ & $\frac{913}{4725}$ & $\frac{386}{4725}$ & $\frac{773}{9450}$ & $\frac{1789}{9450}$ & $\frac{12611}{75600}$ & $\frac{4483}{75600}$ \\
    \hline
    $\frac{47}{3024}$ & $\frac{935}{3024}$ & $\frac{47}{504}$ & $\frac{95}{504}$ & $\frac{341}{4536}$ & $\frac{85}{504}$ & $\frac{1103}{9072}$ & $\frac{85}{3024}$ \\
    \hline
    $\frac{31}{216}$ & $\frac{67}{810}$ & $\frac{191}{720}$ & $\frac{313}{6480}$ & $\frac{547}{6480}$ & $\frac{109}{720}$ & $\frac{59}{405}$ & $\frac{17}{216}$ \\
    \hline
    $\frac{37}{540}$ & $\frac{83}{540}$ & $\frac{79}{432}$ & $\frac{211}{2160}$ & $\frac{209}{2160}$ & $\frac{77}{432}$ & $\frac{41}{270}$ & $\frac{19}{270}$ \\
    \hline
    $\frac{433}{8640}$ & $\frac{2057}{8640}$ & $\frac{23}{192}$ & $\frac{1331}{8640}$ & $\frac{2207}{25920}$ & $\frac{11}{64}$ & $\frac{3469}{25920}$ & $\frac{407}{8640}$ \\
    \hline
    $\frac{107}{1728}$ & $\frac{1481}{5184}$ & $\frac{107}{576}$ & $\frac{955}{5184}$ & $\frac{95}{1728}$ & $\frac{61}{576}$ & $\frac{149}{1728}$ & $\frac{61}{1728}$ \\
    \hline
    0 & $\frac{1}{9}$ & $\frac{4}{9}$ & 0 & 0 & $\frac{2}{9}$ & $\frac{2}{9}$ & 0 \\
    \hline
    $\frac{7}{144}$ & $\frac{5}{48}$ & $\frac{7}{24}$ & $\frac{5}{72}$ & $\frac{7}{72}$ & $\frac{5}{24}$ & $\frac{7}{48}$ & $\frac{5}{144}$ \\
    \hline
    $\frac{8}{81}$ & $\frac{7}{81}$ & $\frac{28}{81}$ & $\frac{2}{81}$ & $\frac{4}{81}$ & $\frac{14}{81}$ & $\frac{14}{81}$ & $\frac{4}{81}$ \\
    \hline
    $\frac{91}{1620}$ & $\frac{71}{648}$ & $\frac{1841}{6480}$ & $\frac{83}{1296}$ & $\frac{581}{6480}$ & $\frac{263}{1296}$ & $\frac{497}{3240}$ & $\frac{13}{324}$ \\
    \hline
    $\frac{16}{405}$ & $\frac{14}{81}$ & $\frac{64}{405}$ & $\frac{11}{81}$ & $\frac{44}{405}$ & $\frac{16}{81}$ & $\frac{56}{405}$ & $\frac{4}{81}$ \\
    \hline
    $\frac{7}{54}$ & $\frac{17}{216}$ & $\frac{17}{54}$ & $\frac{7}{216}$ & $\frac{7}{108}$ & $\frac{17}{108}$ & $\frac{17}{108}$ & $\frac{7}{108}$ \\
    \hline
    $\frac{119}{1080}$ & $\frac{17}{216}$ & $\frac{497}{2160}$ & $\frac{41}{432}$ & $\frac{287}{2160}$ & $\frac{71}{432}$ & $\frac{119}{1080}$ & $\frac{17}{216}$ \\
    \hline
    $\frac{137}{2835}$ & $\frac{9}{56}$ & $\frac{47}{315}$ & $\frac{671}{4536}$ & $\frac{671}{5670}$ & $\frac{47}{252}$ & $\frac{9}{70}$ & $\frac{137}{2268}$ \\
    \hline
    $\frac{1}{112}$ & $\frac{107}{336}$ & $\frac{3}{56}$ & $\frac{41}{168}$ & $\frac{41}{504}$ & $\frac{9}{56}$ & $\frac{107}{1008}$ & $\frac{3}{112}$ \\
    \hline
    $\frac{7}{72}$ & $\frac{5}{54}$ & $\frac{35}{144}$ & $\frac{35}{432}$ & $\frac{49}{432}$ & $\frac{25}{144}$ & $\frac{7}{54}$ & $\frac{5}{72}$ \\
    \hline
    $\frac{4}{81}$ & $\frac{5}{27}$ & $\frac{4}{27}$ & $\frac{10}{81}$ & $\frac{8}{81}$ & $\frac{5}{27}$ & $\frac{4}{27}$ & $\frac{5}{81}$ \\
    \hline
  \end{tabular}
  \caption{Same as Table \ref{probability-diquark-0}. for $q^6s^3$}
  \label{probability-diquark-3}
\end{table}
\end{center}

\begin{center}
\begin{table}
  \begin{tabular}{|c|c|c|c|c|}
    \hline
    $q^5 s^4$ & \multicolumn{4}{|c|}{$i,j=1$-5} \\
    \hline  \hline
    Flavor & $A$ & $S$ & $A$ & $S$   \\
    \hline
    Color  & $A$ & $A$ & $S$ & $S$   \\
    \hline
    Spin  & $A$ & $S$ & $S$ & $A$   \\
    \hline
    $P_{ij}(F$=64,$I$=$\frac{5}{2}$,$S$=$\frac{3}{2}$) & 0 & $\frac{3}{5}$ & 0 & $\frac{2}{5}$ \\
    \hline
    $P_{ij}(F$=$\overline{35}$,$I$=$\frac{5}{2}$,$S$=$\frac{1}{2}$) & 0 & $\frac{3}{5}$ & 0 & $\frac{2}{5}$ \\
    \hline
    $P_{ij}(F$=27,$I$=$\frac{3}{2}$,$S$=$\frac{5}{2}$) & $\frac{1}{20}$ & $\frac{11}{20}$ & $\frac{1}{5}$ & $\frac{1}{5}$ \\
    \hline
    $P_{ij}(F$=$\overline{10}$,$I$=$\frac{3}{2}$,$S$=$\frac{3}{2}$) & $\frac{1}{16}$ & $\frac{25}{48}$ & $\frac{3}{16}$ & $\frac{11}{48}$ \\
    \hline
    $P_{ij}(F$=27,$I$=$\frac{3}{2}$,$S$=$\frac{3}{2}$) & $\frac{43}{560}$ & $\frac{39}{80}$ & $\frac{97}{560}$ & $\frac{21}{80}$ \\
    \hline
    $P_{ij}(F$=64,$I$=$\frac{3}{2}$,$S$=$\frac{3}{2}$) & $\frac{1}{28}$ & $\frac{31}{60}$ & $\frac{3}{14}$ & $\frac{7}{30}$ \\
    \hline
    $P_{ij}(F$=27,$I$=$\frac{3}{2}$,$S$=$\frac{1}{2}$) & $\frac{7}{80}$ & $\frac{39}{80}$ & $\frac{13}{80}$ & $\frac{21}{80}$ \\
    \hline
    $P_{ij}(F$=35,$I$=$\frac{3}{2}$,$S$=$\frac{1}{2}$) & $\frac{1}{10}$ & $\frac{7}{15}$ & $\frac{3}{20}$ & $\frac{17}{60}$ \\
    \hline
    $P_{ij}(F$=$\overline{35}$,$I$=$\frac{3}{2}$,$S$=$\frac{1}{2}$) & $\frac{1}{16}$ & $\frac{119}{240}$ & $\frac{3}{16}$ & $\frac{61}{240}$ \\
    \hline
    $P_{ij}(F$=8,$I$=$\frac{1}{2}$,$S$=$\frac{7}{2}$) & 0 & $\frac{3}{5}$ & $\frac{2}{5}$ & 0 \\
    \hline
    $P_{ij}(F$=8,$I$=$\frac{1}{2}$,$S$=$\frac{5}{2}$) & $\frac{21}{200}$ & $\frac{99}{200}$ & $\frac{59}{200}$ & $\frac{21}{200}$ \\
    \hline
    $P_{ij}(F$=27,$I$=$\frac{1}{2}$,$S$=$\frac{5}{2}$) & $\frac{1}{50}$ & $\frac{29}{50}$ & $\frac{19}{50}$ & $\frac{1}{50}$ \\
    \hline
    $P_{ij}(F$=8,$I$=$\frac{1}{2}$,$S$=$\frac{3}{2}$) & $\frac{21}{200}$ & $\frac{89}{200}$ & $\frac{59}{200}$ & $\frac{31}{200}$ \\
    \hline
    $P_{ij}(F$=10,$I$=$\frac{1}{2}$,$S$=$\frac{3}{2}$) & $\frac{11}{80}$ & $\frac{109}{240}$ & $\frac{21}{80}$ & $\frac{7}{48}$ \\
    \hline
    $P_{ij}(F$=27,$I$=$\frac{1}{2}$,$S$=$\frac{3}{2}$) & $\frac{281}{2800}$ & $\frac{1389}{2800}$ & $\frac{839}{2800}$ & $\frac{291}{2800}$ \\
    \hline
    $P_{ij}(F$=64,$I$=$\frac{1}{2}$,$S$=$\frac{3}{2}$) & $\frac{2}{35}$ & $\frac{53}{105}$ & $\frac{12}{35}$ & $\frac{2}{21}$ \\
    \hline
    $P_{ij}(F$=8,$I$=$\frac{1}{2}$,$S$=$\frac{1}{2}$) & $\frac{7}{40}$ & $\frac{47}{120}$ & $\frac{9}{40}$ & $\frac{5}{24}$ \\
    \hline
    $P_{ij}(F$=27,$I$=$\frac{1}{2}$,$S$=$\frac{1}{2}$) & $\frac{13}{80}$ & $\frac{33}{80}$ & $\frac{19}{80}$ & $\frac{3}{16}$ \\
    \hline
    $P_{ij}(F$=35,$I$=$\frac{1}{2}$,$S$=$\frac{1}{2}$) & $\frac{11}{80}$ & $\frac{101}{240}$ & $\frac{21}{80}$ & $\frac{43}{240}$ \\
    \hline
  \end{tabular}
  \begin{tabular}{|c|c|}
    \hline
    \multicolumn{2}{|c|}{$i,j=6$-9} \\
    \hline  \hline
     ~$S$~ & $S$   \\
    \hline
     $S$ & $A$   \\
    \hline
     $A$ & $S$   \\
    \hline
     $\frac{2}{3}$ & $\frac{1}{3}$  \\
    \hline
     $\frac{2}{3}$ & $\frac{1}{3}$  \\
    \hline
    $\frac{2}{3}$ & $\frac{1}{3}$  \\
    \hline
    $\frac{23}{36}$ & $\frac{13}{36}$  \\
    \hline
    $\frac{17}{28}$ & $\frac{11}{28}$  \\
    \hline
    $\frac{37}{63}$ & $\frac{26}{63}$  \\
    \hline
    $\frac{5}{8}$ & $\frac{3}{8}$  \\
    \hline
    $\frac{11}{18}$ & $\frac{7}{18}$  \\
    \hline
    $\frac{43}{72}$ & $\frac{29}{72}$  \\
    \hline
    $\frac{2}{3}$ & $\frac{1}{3}$  \\
    \hline
    $\frac{2}{3}$ & $\frac{1}{3}$  \\
    \hline
    $\frac{2}{3}$ & $\frac{1}{3}$  \\
    \hline
    $\frac{7}{12}$ & $\frac{5}{12}$  \\
    \hline
    $\frac{47}{72}$ & $\frac{25}{72}$  \\
    \hline
    $\frac{37}{56}$ & $\frac{19}{56}$  \\
    \hline
    $\frac{38}{63}$ & $\frac{25}{63}$  \\
    \hline
    $\frac{11}{18}$ & $\frac{7}{18}$  \\
    \hline
    $\frac{5}{8}$ & $\frac{3}{8}$  \\
    \hline
    $\frac{43}{72}$ & $\frac{29}{71}$  \\
    \hline
  \end{tabular}
  \begin{tabular}{|c|c|c|c|c|c|c|c|c|}
    \hline
    \multicolumn{8}{|c|}{$i=1$-5,$j=6$-9} \\
    \hline  \hline
    $A$ & $S$ & $A$ & $S$  & $A$ & $S$ & $A$ & $S$ \\
    \hline
    $A$ & $A$ & $S$ & $S$ & $S$ & $S$ & $A$ & $A$  \\
    \hline
    $A$ & $S$ & $S$ & $A$ & $A$ & $S$ & $S$ & $A$  \\
    \hline
    $\frac{63}{1600}$ & $\frac{247}{1600}$ & $\frac{189}{800}$ & $\frac{57}{800}$ & $\frac{63}{800}$ & $\frac{171}{800}$ & $\frac{273}{1600}$ & $\frac{57}{1600}$ \\
    \hline
    $\frac{9}{100}$ & $\frac{1}{10}$ & $\frac{27}{100}$ & $\frac{3}{50}$ & $\frac{9}{100}$ & $\frac{9}{50}$ & $\frac{3}{20}$ & $\frac{3}{50}$ \\
    \hline
    $\frac{211}{4800}$ & $\frac{659}{4800}$ & $\frac{23}{80}$ & $\frac{53}{1200}$ & $\frac{67}{1200}$ & $\frac{17}{80}$ & $\frac{901}{4800}$ & $\frac{149}{4800}$ \\
    \hline
    $\frac{25}{256}$ & $\frac{25}{256}$ & $\frac{75}{256}$ & $\frac{11}{256}$ & $\frac{55}{768}$ & $\frac{45}{256}$ & $\frac{125}{768}$ & $\frac{15}{256}$ \\
    \hline
    $\frac{9743}{134400}$ & $\frac{17807}{134400}$ & $\frac{6961}{26880}$ & $\frac{6581}{134400}$ & $\frac{1237}{19200}$ & $\frac{737}{3840}$ & $\frac{3439}{19200}$ & $\frac{991}{19200}$ \\
    \hline
    $\frac{19}{700}$ & $\frac{533}{2100}$ & $\frac{57}{350}$ & $\frac{143}{1050}$ & $\frac{11}{150}$ & $\frac{9}{50}$ & $\frac{41}{300}$ & $\frac{3}{100}$ \\
    \hline
    $\frac{349}{3840}$ & $\frac{2189}{19200}$ & $\frac{4619}{19200}$ & $\frac{259}{3840}$ & $\frac{341}{3840}$ & $\frac{3421}{19200}$ & $\frac{2971}{19200}$ & $\frac{251}{3840}$ \\
    \hline
    $\frac{11}{200}$ & $\frac{43}{240}$ & $\frac{57}{320}$ & $\frac{499}{4800}$ & $\frac{421}{4800}$ & $\frac{63}{320}$ & $\frac{37}{240}$ & $\frac{9}{200}$ \\
    \hline
    $\frac{469}{6400}$ & $\frac{269}{1280}$ & $\frac{1407}{6400}$ & $\frac{851}{6400}$ & $\frac{1327}{19200}$ & $\frac{873}{6400}$ & $\frac{433}{3840}$ & $\frac{291}{6400}$ \\
    \hline
    $\frac{3}{64}$ & $\frac{39}{320}$ & $\frac{11}{32}$ & $\frac{3}{160}$ & $\frac{1}{32}$ & $\frac{33}{160}$ & $\frac{13}{64}$ & $\frac{9}{320}$ \\
    \hline
    $\frac{11}{160}$ & $\frac{87}{800}$ & $\frac{103}{320}$ & $\frac{51}{1600}$ & $\frac{17}{320}$ & $\frac{309}{1600}$ & $\frac{29}{160}$ & $\frac{33}{800}$ \\
    \hline
    $\frac{1}{24}$ & $\frac{389}{2400}$ & $\frac{7}{32}$ & $\frac{59}{600}$ & $\frac{11}{120}$ & $\frac{153}{800}$ & $\frac{71}{480}$ & $\frac{29}{600}$ \\
    \hline
    $\frac{69}{640}$ & $\frac{333}{3200}$ & $\frac{181}{640}$ & $\frac{117}{3200}$ & $\frac{39}{640}$ & $\frac{543}{3200}$ & $\frac{111}{640}$ & $\frac{207}{3200}$ \\
    \hline
    $\frac{89}{1280}$ & $\frac{169}{1280}$ & $\frac{309}{1280}$ & $\frac{101}{1280}$ & $\frac{313}{3840}$ & $\frac{243}{1280}$ & $\frac{121}{768}$ & $\frac{63}{1280}$ \\
    \hline
    $\frac{19}{336}$ & $\frac{5191}{33600}$ & $\frac{137}{672}$ & $\frac{3559}{33600}$ & $\frac{691}{6720}$ & $\frac{193}{1050}$ & $\frac{919}{6720}$ & $\frac{937}{16800}$ \\
    \hline
    $\frac{53}{2240}$ & $\frac{2039}{6720}$ & $\frac{159}{1120}$ & $\frac{713}{3360}$ & $\frac{239}{3360}$ & $\frac{153}{1120}$ & $\frac{593}{6720}$ & $\frac{51}{2240}$ \\
    \hline
    $\frac{7}{64}$ & $\frac{31}{320}$ & $\frac{9}{32}$ & $\frac{7}{160}$ & $\frac{7}{96}$ & $\frac{27}{160}$ & $\frac{31}{192}$ & $\frac{21}{320}$ \\
    \hline
    $\frac{49}{768}$ & $\frac{611}{3840}$ & $\frac{151}{768}$ & $\frac{389}{3840}$ & $\frac{71}{768}$ & $\frac{709}{3840}$ & $\frac{113}{768}$ & $\frac{211}{3840}$ \\
    \hline
    $\frac{5}{128}$ & $\frac{433}{1920}$ & $\frac{87}{640}$ & $\frac{287}{1920}$ & $\frac{173}{1920}$ & $\frac{117}{640}$ & $\frac{259}{1920}$ & $\frac{27}{640}$ \\
    \hline
  \end{tabular}
  \caption{Same as Table \ref{probability-diquark-0}. for $q^5s^4$}
  \label{probability-diquark-4}
\end{table}
\end{center}

\begin{center}
\begin{table}
  \begin{tabular}{|c|c|c|c|c|}
    \hline
    $q^4 s^5$ & \multicolumn{4}{|c|}{$i,j=1$-4} \\
    \hline  \hline
    Flavor & $A$ & $S$ & $A$ & $S$   \\
    \hline
    Color  & $A$ & $A$ & $S$ & $S$   \\
    \hline
    Spin  & $A$ & $S$ & $S$ & $A$   \\
    \hline
    $P_{ij}(F$=64,$I$=2,$S$=$\frac{3}{2}$) & 0 & $\frac{2}{3}$ & 0 & $\frac{1}{3}$ \\
    \hline
    $P_{ij}(F$=$\overline{35}$,$I$=2,$S$=$\frac{1}{2}$) & 0 & $\frac{2}{3}$ & 0 & $\frac{1}{3}$ \\
    \hline
    $P_{ij}(F$=27,$I$=1,$S$=$\frac{5}{2}$) & 0 & $\frac{2}{3}$ & $\frac{1}{3}$ & 0 \\
    \hline
    $P_{ij}(F$=27,$I$=1,$S$=$\frac{3}{2}$) & $\frac{5}{42}$ & $\frac{23}{42}$ & $\frac{3}{14}$ & $\frac{5}{42}$ \\
    \hline
    $P_{ij}(F$=64,$I$=1,$S$=$\frac{3}{2}$) & $\frac{1}{21}$ & $\frac{13}{21}$ & $\frac{2}{7}$ & $\frac{1}{21}$ \\
    \hline
    $P_{ij}(F$=27,$I$=1,$S$=$\frac{1}{2}$) & $\frac{5}{24}$ & $\frac{11}{24}$ & $\frac{1}{8}$ & $\frac{5}{24}$ \\
    \hline
    $P_{ij}(F$=35,$I$=1,$S$=$\frac{1}{2}$) & $\frac{5}{24}$ & $\frac{11}{24}$ & $\frac{1}{8}$ & $\frac{5}{24}$ \\
    \hline
    $P_{ij}(F$=10,$I$=0,$S$=$\frac{3}{2}$) & $\frac{1}{4}$ & $\frac{5}{12}$ & $\frac{1}{4}$ & $\frac{1}{12}$ \\
    \hline
    $P_{ij}(F$=35,$I$=0,$S$=$\frac{1}{2}$) & $\frac{1}{4}$ & $\frac{5}{12}$ & $\frac{1}{4}$ & $\frac{1}{12}$ \\
    \hline
  \end{tabular}
  \begin{tabular}{|c|c|}
    \hline
    \multicolumn{2}{|c|}{$i,j=5$-9} \\
    \hline  \hline
     ~$S$~ & $S$   \\
    \hline
     $S$ & $A$   \\
    \hline
     $A$ & $S$   \\
    \hline
     $\frac{3}{5}$ & $\frac{2}{5}$  \\
    \hline
     $\frac{3}{5}$ & $\frac{2}{5}$  \\
    \hline
     $\frac{3}{5}$ & $\frac{2}{5}$  \\
    \hline
     $\frac{3}{5}$ & $\frac{2}{5}$  \\
    \hline
     $\frac{3}{5}$ & $\frac{2}{5}$  \\
    \hline
     $\frac{3}{5}$ & $\frac{2}{5}$  \\
    \hline
     $\frac{3}{5}$ & $\frac{2}{5}$  \\
    \hline
     $\frac{3}{5}$ & $\frac{2}{5}$  \\
    \hline
     $\frac{3}{5}$ & $\frac{2}{5}$  \\
    \hline
  \end{tabular}
  \begin{tabular}{|c|c|c|c|c|c|c|c|c|}
    \hline
    \multicolumn{8}{|c|}{$i=1$-4,$j=5$-9} \\
    \hline  \hline
    $A$ & $S$ & $A$ & $S$  & $A$ & $S$ & $A$ & $S$ \\
    \hline
    $A$ & $A$ & $S$ & $S$ & $S$ & $S$ & $A$ & $A$  \\
    \hline
    $A$ & $S$ & $S$ & $A$ & $A$ & $S$ & $S$ & $A$  \\
    \hline
    $\frac{63}{1600}$ & $\frac{247}{1600}$ & $\frac{189}{800}$ & $\frac{57}{800}$ & $\frac{63}{800}$ & $\frac{171}{800}$ & $\frac{273}{1600}$ & $\frac{57}{1600}$ \\
    \hline
    $\frac{9}{100}$ & $\frac{1}{10}$ & $\frac{27}{100}$ & $\frac{3}{50}$ & $\frac{9}{100}$ & $\frac{9}{50}$ & $\frac{3}{20}$ & $\frac{3}{50}$ \\
    \hline
    $\frac{3}{50}$ & $\frac{3}{25}$ & $\frac{3}{10}$ & $\frac{1}{25}$ & $\frac{3}{50}$ & $\frac{1}{5}$ & $\frac{9}{50}$ & $\frac{1}{25}$ \\
    \hline
    $\frac{213}{2800}$ & $\frac{153}{1400}$ & $\frac{159}{560}$ & $\frac{71}{1400}$ & $\frac{213}{2800}$ & $\frac{53}{280}$ & $\frac{459}{2800}$ & $\frac{71}{1400}$ \\
    \hline
    $\frac{387}{11200}$ & $\frac{2619}{11200}$ & $\frac{1161}{5600}$ & $\frac{177}{1120}$ & $\frac{87}{1120}$ & $\frac{879}{5600}$ & $\frac{1181}{11200}$ & $\frac{293}{11200}$ \\
    \hline
    $\frac{33}{400}$ & $\frac{21}{200}$ & $\frac{111}{400}$ & $\frac{11}{200}$ & $\frac{33}{400}$ & $\frac{37}{200}$ & $\frac{63}{400}$ & $\frac{11}{200}$ \\
    \hline
    $\frac{3}{64}$ & $\frac{51}{320}$ & $\frac{333}{1600}$ & $\frac{153}{1600}$ & $\frac{147}{1600}$ & $\frac{327}{1600}$ & $\frac{49}{320}$ & $\frac{13}{320}$ \\
    \hline
    $\frac{27}{400}$ & $\frac{23}{200}$ & $\frac{117}{400}$ & $\frac{9}{200}$ & $\frac{27}{400}$ & $\frac{39}{200}$ & $\frac{69}{400}$ & $\frac{9}{200}$ \\
    \hline
    $\frac{27}{800}$ & $\frac{143}{800}$ & $\frac{27}{160}$ & $\frac{99}{800}$ & $\frac{81}{800}$ & $\frac{33}{160}$ & $\frac{117}{800}$ & $\frac{33}{800}$ \\
    \hline
  \end{tabular}
  \caption{Same as Table \ref{probability-diquark-0}. for $q^4s^5$}
  \label{probability-diquark-5}
\end{table}
\end{center}

\begin{center}
\begin{table}
  \begin{tabular}{|c|c|c|c|c|}
    \hline
    $q^3 s^6$ & \multicolumn{4}{|c|}{$i,j=1$-3} \\
    \hline  \hline
    Flavor & $A$ & $S$ & $A$ & $S$   \\
    \hline
    Color  & $A$ & $A$ & $S$ & $S$   \\
    \hline
    Spin  & $A$ & $S$ & $S$ & $A$   \\
    \hline
    $P_{ij}(F$=64,$I$=$\frac{3}{2}$,$S$=$\frac{3}{2}$) & 0 & 1 & 0 & 0 \\
    \hline
    $P_{ij}(F$=35,$I$=$\frac{1}{2}$,$S$=$\frac{1}{2}$) & $\frac{1}{2}$ & $\frac{1}{2}$ & 0 & 0 \\
    \hline
  \end{tabular}
  \begin{tabular}{|c|c|}
    \hline
    \multicolumn{2}{|c|}{$i,j=4$-9} \\
    \hline  \hline
     ~$S$~ & $S$   \\
    \hline
     $S$ & $A$   \\
    \hline
     $A$ & $S$   \\
    \hline
     $\frac{3}{5}$ & $\frac{2}{5}$  \\
    \hline
     $\frac{3}{5}$ & $\frac{2}{5}$  \\
    \hline
  \end{tabular}
  \begin{tabular}{|c|c|c|c|c|c|c|c|c|}
    \hline
    \multicolumn{8}{|c|}{$i=1$-3,$j=4$-9} \\
    \hline  \hline
    $A$ & $S$ & $A$ & $S$  & $A$ & $S$ & $A$ & $S$ \\
    \hline
    $A$ & $A$ & $S$ & $S$ & $S$ & $S$ & $A$ & $A$  \\
    \hline
    $A$ & $S$ & $S$ & $A$ & $A$ & $S$ & $S$ & $A$  \\
    \hline
    $\frac{7}{144}$ & $\frac{5}{48}$ & $\frac{7}{24}$ & $\frac{5}{72}$ & $\frac{7}{72}$ & $\frac{5}{24}$ & $\frac{7}{48}$ & $\frac{5}{144}$ \\
    \hline
    $\frac{7}{144}$ & $\frac{5}{48}$ & $\frac{7}{24}$ & $\frac{5}{72}$ & $\frac{7}{72}$ & $\frac{5}{24}$ & $\frac{7}{48}$ & $\frac{5}{144}$ \\
    \hline
  \end{tabular}
  \caption{Same as Table \ref{probability-diquark-0}. for $q^3s^6$}
  \label{probability-diquark-6}
\end{table}
\end{center}

\end{widetext}

\section{Summary}
\label{Summary}
In this work, we calculate the matrix elements of color-spin interaction of a tribaryon in the flavor SU(3) broken case. To calculate this, we construct the flavor$\otimes$color$\otimes$spin wave function of a tribaryon to satisfy the Pauli principle. Additionally, we analyze the diquark structure of a tribaryon by using the symmetric and antisymmetric basis set of each Young tableau of $S_9$ symmetric group. By looking at the largest diquark components for antisymmetric color and spin states, we can find the most attractive channels of a tribaryon. Furthermore, The diquark structure table can be useful in analyzing the flavor-spin interactions as well as the color-spin interactions of a tribaryon.

\section*{Acknowledgments}
The work by SHL was supported by Samsung Science and Technology Foundation under Project Number SSTF-BA1901-04.  This work by AP was supported by the Korea National Research
Foundation under the grant number 2018R1D1A1B07043234.

\begin{appendix}

\section{Orbital state of a tribaryon}
\label{orbital}
In this section, we present the possible orbital state of a tribaryon considering it as a combined state of three s-wave baryons. For a baryon, we can represent its orbital state using Young diagram as [3] since the orbital part is totally symmetric. it is also well known that there are in total four possible orbital state in a two-baryon system as follows.
\begin{align}
  [3]\times[3]=[6]+[51]+[42]+[33].
\end{align}
\label{orbital-baryon}

By further applying the outer product of an orbital state of a baryon, we can get the following possible orbital state of a tribaryon.

\begin{align}
  [3]&\times[3]\times[3] \nonumber\\
  =&[6]\times[3]+[51]\times[3]+[42]\times[3]+[33]\times[3] \nonumber\\
  =&[9]+[81]_{(m=2)}+[72]_{(m=3)}+[63]_{(m=4)}+[54]_{(m=2)}+ \nonumber\\
  &[711]+[621]_{(m=2)}+[531]_{(m=3)}+[522]+[432]_{(m=2)}+ \nonumber\\
  &[333],
\label{orbital-tribaryon}
\end{align}
where $m$ is the multiplicity of the corresponding state. We can also represent the orbital state of a tribaryon in terms of baryon configuration. There are four possible representations as follows.
\begin{align}
  \Ket{O_1}&=
  \Ket{\begin{tabular}{|c|c|c|}
  \cline{1-3}
  $B_1$ & $B_2$ & $B_3$  \\
  \cline{1-3}
  \end{tabular}},\quad
  \Ket{O_2}=
  \Ket{\begin{tabular}{|c|c|}
  \cline{1-2}
  $B_1$ & $B_2$ \\
  \cline{1-2}
  $B_3$ \\
  \cline{1-1}
  \end{tabular}}, \nonumber\\
  \Ket{O_3}&=
  \Ket{\begin{tabular}{|c|c|}
  \cline{1-2}
  $B_1$ & $B_3$ \\
  \cline{1-2}
  $B_2$ \\
  \cline{1-1}
  \end{tabular}}, \quad
  \Ket{O_4}=
  \Ket{\begin{tabular}{|c|}
  \cline{1-1}
  $B_1$ \\
  \cline{1-1}
  $B_2$ \\
  \cline{1-1}
  $B_3$ \\
  \cline{1-1}
  \end{tabular}}
\label{orbital-baryon-basis}
\end{align}
where
$\begin{tabular}{|c|}
\cline{1-1}
$B_i$ \\
\cline{1-1}
\end{tabular}$
represent an orbital state of the $i$-th baryon. Now, $\Ket{O_1}$ and $\Ket{O_4}$ states are symmetric and antisymmetric under the exchange among three baryons, respectively.

On the other hand, $\Ket{O_2}$ and $\Ket{O_3}$ states have mixed symmetric property. By constructing the orbital state using the above representation, we can connect this representation to Eq.(\ref{orbital-tribaryon}). For [9], it is obvious that it corresponds to $\Ket{O_1}$. For [8,1], let's consider the following orbital basis set for [8,1].\\
\\
$\Ket{O_1^{[8,1]}}=
  \Ket{\begin{tabular}{|c|c|c|c|c|c|c|c|}
  \cline{1-8}
  1 & 2 & 3 & 4 & 5 & 6 & 7 & 8  \\
  \cline{1-8}
  9 \\
  \cline{1-1}
  \end{tabular}}_O$,\\
$\Ket{O_2^{[8,1]}}=
  \Ket{\begin{tabular}{|c|c|c|c|c|c|c|c|}
  \cline{1-8}
  1 & 2 & 3 & 4 & 5 & 6 & 7 & 9  \\
  \cline{1-8}
  8 \\
  \cline{1-1}
  \end{tabular}}_O$,\\
$\Ket{O_3^{[8,1]}}=
  \Ket{\begin{tabular}{|c|c|c|c|c|c|c|c|}
  \cline{1-8}
  1 & 2 & 3 & 4 & 5 & 6 & 8 & 9  \\
  \cline{1-8}
  7 \\
  \cline{1-1}
  \end{tabular}}_O$,\\
$\Ket{O_4^{[8,1]}}=
  \Ket{\begin{tabular}{|c|c|c|c|c|c|c|c|}
  \cline{1-8}
  1 & 2 & 3 & 4 & 5 & 7 & 8 & 9  \\
  \cline{1-8}
  6 \\
  \cline{1-1}
  \end{tabular}}_O$,\\
$\Ket{O_5^{[8,1]}}=
  \Ket{\begin{tabular}{|c|c|c|c|c|c|c|c|}
  \cline{1-8}
  1 & 2 & 3 & 4 & 6 & 7 & 8 & 9  \\
  \cline{1-8}
  5 \\
  \cline{1-1}
  \end{tabular}}_O$,\\
$\Ket{O_6^{[8,1]}}=
  \Ket{\begin{tabular}{|c|c|c|c|c|c|c|c|}
  \cline{1-8}
  1 & 2 & 3 &  5 & 6 & 7 & 8 & 9  \\
  \cline{1-8}
  4 \\
  \cline{1-1}
  \end{tabular}}_O$,\\
$\Ket{O_7^{[8,1]}}=
  \Ket{\begin{tabular}{|c|c|c|c|c|c|c|c|}
  \cline{1-8}
  1 & 2 & 4 & 5 & 6 & 7 & 8 & 9  \\
  \cline{1-8}
  3 \\
  \cline{1-1}
  \end{tabular}}_O$,\\
$\Ket{O_8^{[8,1]}}=
  \Ket{\begin{tabular}{|c|c|c|c|c|c|c|c|}
  \cline{1-8}
  1 & 3 & 4 & 5 & 6 & 7 & 8 & 9  \\
  \cline{1-8}
  2 \\
  \cline{1-1}
  \end{tabular}}_O$.\\

Then we can construct $\Ket{O_2}$ and $\Ket{O_3}$ in Eq.(\ref{orbital-baryon-basis}) using [8,1] basis as follows.\\
\begin{align}
  \Ket{O_2} &= \frac{1}{2}\Ket{O_1^{[8,1]}}+\frac{3}{2\sqrt{7}}\Ket{O_2^{[8,1]}}+\frac{\sqrt{3}}{\sqrt{7}}\Ket{O_3^{[8,1]}} \nonumber\\
  \Ket{O_3} &= \frac{1}{\sqrt{5}}\Ket{O_4^{[8,1]}}+\frac{\sqrt{3}}{\sqrt{10}}\Ket{O_5^{[8,1]}}+\frac{1}{\sqrt{2}}\Ket{O_6^{[8,1]}}.
\end{align}

We can check that $\Ket{O_2}$ is symmetric under (14)(25)(36) and $\Ket{O_3}$ is antisymmetric under (14)(25)(36). Also, the above states satisfy the other symmetric properties which [2,1] Young diagram have. Using similar method, we can find the relation between the two representations with other Young diagrams.
The open circles in table  \ref{orbital-table}, show the non-vanishing  overlap between the two representations.

\begin{center}
\begin{table}[htbp]
\begin{tabular}{c|c|c|c|c}
  \hline
  & $O_1$ & $O_2$ & $O_3$ & $O_4$ \\
  \hline
  [9] & $\bigcirc$ & & & \\
  \hline
  [81] & & $\bigcirc$ & $\bigcirc$ & \\
  \hline
  [72] & $\bigcirc$ & $\bigcirc$ & $\bigcirc$ & \\
  \hline
  [63] & $\bigcirc$ & $\bigcirc$ & $\bigcirc$ & $\bigcirc$ \\
  \hline
  [54] & & $\bigcirc$ & $\bigcirc$ & \\
  \hline
  [711] & & & & $\bigcirc$ \\
  \hline
  [621] & & $\bigcirc$ & $\bigcirc$ & \\
  \hline
  [531] & & $\bigcirc$ & $\bigcirc$ & $\bigcirc$ \\
  \hline
  [522] & $\bigcirc$ & & & \\
  \hline
  [432] & & $\bigcirc$ & $\bigcirc$ & \\
  \hline
  [333] & & & & $\bigcirc$ \\
  \hline
\end{tabular}
\caption{The relation between three baryons and nine quarks configurations. $\bigcirc$ represents that there is a connection between the two configurations.}
\label{orbital-table}
\end{table}
\end{center}

\end{appendix}

\end{document}